\documentclass[10pt,aps,prb,showpacs,floatfix,floats,superscriptaddress,twocolumn]{revtex4-1}
\usepackage{graphicx,latexsym}
\usepackage{dcolumn}
\usepackage{amssymb,amsmath,bm}
\usepackage{color}
\usepackage[normalem]{ulem}

\begin{document}

\title{Electroluminescence caused by the transport of interacting electrons\\ through
       parallel quantum dots in a photon cavity}

\author{Vidar Gudmundsson}
\email{vidar@hi.is}
\affiliation{Science Institute, University of Iceland, Dunhaga 3, IS-107 Reykjavik, Iceland}
\author{Nzar Rauf Abdullah}
\affiliation{Physics Department, College of Science, 
             University of Sulaimani, Kurdistan Region, Iraq}
\author{Anna Sitek}
\affiliation{Science Institute, University of Iceland, Dunhaga 3, IS-107 Reykjavik, Iceland}
\affiliation{Department of Theoretical Physics, Wroc{\l}aw University of Science and Technology, 50-370 Wroc{\l}aw, Poland}
\author{Hsi-Sheng Goan}
\email{goan@phys.ntu.edu.tw}
\affiliation{Department of Physics and Center for Theoretical Sciences, National Taiwan University, 
             Taipei 10617, Taiwan}
\affiliation{Center for Quantum Science and Engineering, 
             National Taiwan University, Taipei 10617, Taiwan}
\author{Chi-Shung Tang}
\email{cstang@nuu.edu.tw}
\affiliation{Department of Mechanical Engineering, National United University, Miaoli 36003, Taiwan}
\author{Andrei Manolescu}
\email{manoles@ru.is}
\affiliation{School of Science and Engineering, Reykjavik University, Menntavegur 
             1, IS-101 Reykjavik, Iceland}

%

\begin{abstract}
We show that a Rabi-splitting of the states of strongly interacting electrons in parallel quantum dots 
embedded in a short quantum wire placed in a photon cavity can be produced by either the para- or the 
dia-magnetic electron-photon interactions when the geometry of the system is properly accounted for and 
the photon field is tuned close to a resonance with the electron system. We use these two resonances to explore
the electroluminescence caused by the transport of electrons through the one- and two-electron ground
states of the system and their corresponding conventional and vacuum electroluminescense as the central system is opened
up by coupling it to external leads acting as electron reservoirs. 
Our analysis indicates that high-order electron-photon processes are necessary to adequately 
construct the cavity-photon dressed electron states needed to describe both types of electroluminescence.

\end{abstract}

\maketitle
%
%

\section{Introduction}

Since the discovery of many-body quantum mechanics and quantum electrodynamics, vacuum 
effects have intrigued researchers. Low-dimensional solid-state systems offer 
a good platform to measure some of these vacuum effects due to their tunable
parameters.\cite{PhysRevLett.69.3314,Ciuti05:115303,PhysRevLett.90.116401} 
In particular, nanoelectronic systems coupled to external 
leads.
The vacuum effects can be amplified in experimental systems by enhancing the
electron-photon interaction through microcavities used to bring the photon
frequency into a resonance with a particular electron transition, 
a Rabi resonance.\cite{Bishop09:105} 
An ultra-strong light-matter coupling was achieved in systems of atoms in 
metal cavities in the late 1980s,\cite{PhysRevLett.58.353,RevModPhys.73.565} and in solid 
state systems few years later.\cite{PhysRevLett.69.3314}    
The parameter commonly used to characterize the strength of the light-matter interaction in 
cavity quantum electrodynamics is the normalized coupling, $\eta = \Omega_R/\omega$,
where $\Omega_R$ is the Rabi frequency of two electronic states brought close to a resonance with the 
frequency of the cavity photon $\omega$.\cite{PhysRevLett.116.113601} For $\eta\ll 1$ the 
one-electron ground state has only a small contribution of states with finite
photon component when the Hamiltonian of the system is diagonalized.   
In the ultra-strong regime when $\eta\approx 1$ the expectation value of the photon operator
in the one-electron ground state is large. An electron entering the system from the
left lead has a finite probability to exit the system, from the ground state,
into the right lead leaving behind a photon in the central system. A current through the electronic system 
in the cavity can thus lead to a Ground State Electroluminescence 
(GSE).\cite{PhysRevLett.116.113601} 
If the lifetime of the cavity photons is longer than the lifetime of the electrons in the central 
system the accumulation of cavity photons can lead to the occupation of high-lying 
photon-dressed electron states well above the bias window defined
by the chemical potentials of the external leads.\cite{PhysRevLett.116.113601}    

A related dynamical phenomena termed, {\lq\lq}extra cavity quantum vacuum radiation{\rq\rq}, 
has been predicted for the case of a fast modulated Rabi frequency in a two-level system 
coupled both resonantly and anti-resonantly to a single photon mode in a cavity without 
a coupling to external leads.\cite{PhysRevA.80.053810}

Here, we investigate how the geometry of the parallel double quantum dots and the polarization 
of the cavity-photon field in a 3D rectangular cavity can be used to study the conventional electroluminescence 
and the corresponding GSE of either the para- or the diamagnetic part of the electron-photon interaction for 
the one- or the Coulomb interacting two-electron ground state of the central system. 
Recently, a gate defined double-dot system in a GaAs heterostructure, but in a different kind of a 
cavity, a planar SQUID-array resonator, has been brought into the strong light-material 
coupling regime.\cite{PhysRevX.7.011030} The threshold dynamics of masing has been studied
in gate defined InAs double quantum dots,\cite{2017arXiv170401961L} and quantum dots in a planar 
microwave cavity have both been coupled to superconducting and fermionic external leads, 
with the transport current into and through the system showing high sensitivity to a low number 
of photons in the respective cavities.\cite{Delbecq11:01,PhysRevX.6.021014}

In order to accomplish our analysis we use a Markovian master equation derived from a non-Markovian
master equation without invoking the rotating wave approximation that would only select the resonant terms
of the electron-photon interaction.\cite{2016arXiv161003223J} As we use a configuration
interaction (CI) approach for the electron-electron and the electron-photon interactions,
we are not limited to observation of the effects in areas of the parameter space accessible 
by low order perturbation calculations. 
In a multi-level system the Rabi splitting of a particular set of states is influenced by
the properties of the near lying and higher energy states of the system. The polarizability of the charge 
distribution in the system depends on its geometry and the interactions present in the system. 
As a result the normalized coupling $\eta$ may not be easily tuned into the ultra strong regime,
but even so, vacuum radiation effects can be observed for lower values of $\eta$.  

Traditionally, approaches to nonequilibrium transport have more commonly been built on
Green functions.\cite{StefanucciBook:2013} Recently, Hagenm{\"u}ller et al.\ used a Green functions formalism to
study the cavity-enhanced transport of charge through a non-interacting one-dimensional
system coupled to external leads. In a supplemental-material section they compare their
formalism to a Markovian master equation.\cite{2017arXiv170300803H} An advantage of the
Green functions is that they can conveniently be used for non-adiabatic conditions,\cite{Moldoveanu07:0706.0968}
but a disadvantage is that many-body Coulomb effects -- which can be naturally incorporated in
a master equation\cite{Moldoveanu10:155442} -- can not be easily included without additional approximations.

\section{The model}
\begin{figure}[htb]
      \includegraphics[width=0.45\textwidth]{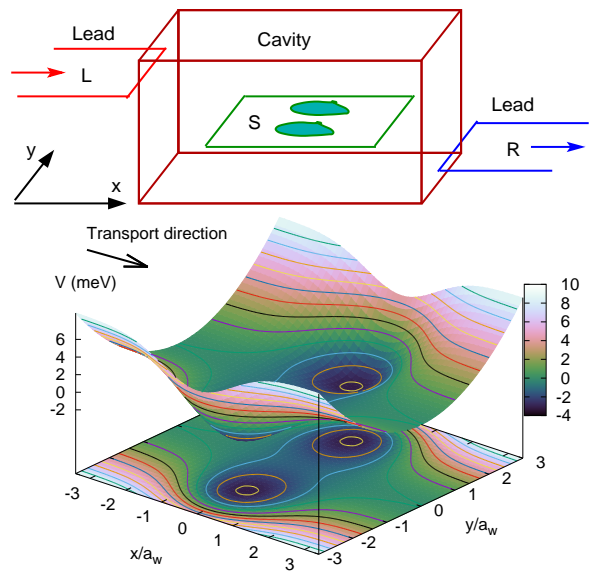}
      \caption{(Top) Schematic of the double parallel dots enbedded in a short quantum wire,
               the central system (S),
               placed in a 3D photon cavity coupled to the left (L) and right (R) leads. 
               The color and the hight of the leads indicates their chemical potentials.
               (Bottom) The potential $V(x,y)$ defining the short quantum wire with the two embedded parallel
               quantum dots. $a_w=23.8$ nm. The arrow defines the transport direction, the $x$-direction.}
     \label{V_2QD}
\end{figure}

We model a two-dimensional electron system in a short quantum wire of length $L_x=150$ nm with two embedded 
parallel quantum dots. The short wire is parabolically confined in the $y$-direction with characteristic
confinement energy $\hbar\Omega_0=2.0$ meV, and has a hard wall confinement at the ends in the $x$-direction,
$\pm L_x/2$, that
will be the direction of electron transport to be described below. The short quantum wire is placed in the 
center of a rectangular photon cavity. The cavity and the two-dimensional electron system will be referred
to as the ``central system'', as it will be coupled to two external leads acting as electron reservoirs
below. The potential describing the closed electronic system can be expressed as
\begin{align}
      V(x,y) =& \left[\frac{1}{2}m^*\Omega_0^2y^2 +eV_g\right.\nonumber\\
             +& \left. V_d\sum_{i=1}^2\exp{\left\{-(\beta x)^2-\beta^2(y-d_i)^2\right\}} \right]\nonumber\\
             \times&\theta\left(\frac{L_x}{2}-x\right)\theta\left(\frac{L_x}{2}+x\right),
\label{Potential}
\end{align}
with $V_d = -6.5$ meV, $\beta = 0.03$ nm$^{-1}$, $d_1=-50$ nm, $d_2=+50$ nm, and $\theta$ is the Heaviside 
step function. $V_g$ is the electrostatic plunger-gate voltage used to place the many-body energy levels
of the central system with respect to the chemical potentials of the external leads to be introduced below.
A schematic of the system, and the shape of the potential (\ref{Potential}) are displayed in Fig.\ \ref{V_2QD}.

The many-body Hamiltonian of the central system in terms of the field operators is
\begin{align}
      H_\mathrm{S} =& \int d^2r \psi^\dagger (\mathbf{r})\left\{\frac{\pi^2}{2m^*}+
        V(\mathbf{r})\right\}\psi (\mathbf{r})
        + H_\mathrm{EM} + H_\mathrm{Coul}\nonumber\\ 
        +H_{\mathrm{Z}}&-\frac{1}{c}\int d^2r\;\mathbf{j}(\mathbf{r})\cdot\mathbf{A}_\gamma
        -\frac{e}{2m^*c^2}\int d^2r\;\rho(\mathbf{r}) A_\gamma^2,
\label{Hclosed}
\end{align}
where 
\begin{equation}
      {\bm{\pi}}=\left(\mathbf{p}+\frac{e}{c}\mathbf{A}_{\mathrm{ext}}\right).
\end{equation}
We assume GaAs parameters for the electron system with $\kappa_e =12.4$, $m^*=0.067m_e$, and $g^*=-0.44$.
The cavity-photon field operator, in terms of the creation and annihilation operators, 
in a stacked notation is expressed as
\begin{equation}
      \mathbf{A}_\gamma (\mathbf{r})=\left({\hat{\mathbf{e}}_x \atop \hat{\mathbf{e}}_y}\right)
      {\cal A}\left\{a+a^{\dagger}\right\}
      \left({\cos{\left(\frac{\pi y}{a_\mathrm{c}}\right)}\atop\cos{\left(\frac{\pi x}{a_\mathrm{c}}\right)}} \right)
      \cos{\left(\frac{\pi z}{d_\mathrm{c}}\right)},
\label{Cav-A}
\end{equation}
for the TE$_{011}$ ($x$-polarization) and TE$_{101}$ ($y$-polarization) modes, respectively.
The strength of the vector potential, ${\cal A}$,
is determined by the coupling constant $g_\mathrm{EM} = e{\cal A}\Omega_wa_w/c$, 
leaving a dimensionless polarization tensor
\begin{equation}
      g_{ij}^k = \frac{a_w}{2\hbar}\left\{\langle i|\hat{\mathbf{e}}_k\cdot\bm{\pi}|j\rangle + \mathrm{h.c.}\right\},  
\end{equation}
where $|i\rangle$ and $|j\rangle$ are single-electron states of the short two-dimensional quantum wire,
$k=x,\,\mbox{or}\,\, y$. 
(Latin indices are used for the single-electron states, and Greek for the 
many body states to be described below).
In the far-infrared regime, with photon energy in the range 0.7-2.0 meV, the characteristic lengths of the 
photon cavity are much larger than the size of the electronic system, $L_x$, and we approximate the cosines
with 1 in the center of the cavity.
The Hamiltonian of the single cavity-photon mode is $H_\mathrm{EM}=\hbar\omega a^\dagger a$,
and for the Coulomb interaction we use
\begin{equation}
      H_\mathrm{Coul} = \frac{1}{2}\int d^2rd^2r' \psi^\dagger (\mathbf{r}) \psi^\dagger (\mathbf{r}')
                        V_{\mathrm{Coul}}(\mathbf{r}-\mathbf{r}')\psi (\mathbf{r}') \psi (\mathbf{r}) ,
\end{equation}
with a spatial dependent Coulomb kernel\cite{Gudmundsson:2013.305} 
\begin{equation}
      V_{\mathrm{Coul}}(\mathbf{r}-\mathbf{r}') = \frac{e^2}{\kappa_e\sqrt{|\mathbf{r}-\mathbf{r}'|^2+\eta_c^2}},
\end{equation}
and a small regularizing parameter $\eta_c/a_w=3\times 10^{-7}$.   

The second term in the second line of Hamiltonian of the central system ({\ref{Hclosed}}) is the
paramagnetic electron-photon interaction, and the third term is the diamagnetic electron-photon interaction
proportional to the integral of $A_\gamma^2$ and the electron charge density $\rho$. 
The external homogeneous magnetic field $\mathbf{B}=\mathbf{\nabla}\times\mathbf{A}_\mathrm{ext}$
is set to 0.1 T in order to break all spin degeneracies. The corresponding Zeeman splitting is 
described by $H_\mathrm{Z}$. The charge and charge-current density operators are 
\begin{equation}
      \rho       = -e\psi^\dagger\psi, \quad
      \mathbf{j} = -\frac{e}{2m^*}\left\{\psi^\dagger\left({\bm{\pi}}\psi\right)
                 +\left({\bm{\pi}}^*\psi^\dagger\right)\psi\right\}.
\end{equation}
The external magnetic field $B$ and the parabolic confinement energy of the 
central system $\hbar\Omega_0$ lead together, with the cyclotron frequency 
$\omega_c=((eB)/(m^*c))$, to an effective characteristic frequency 
$\Omega_w=({\omega_c^2+\Omega_0^2})^{1/2}$ and an effective magnetic length $a_w=(\hbar /(m^*\Omega_w))^{1/2}$.
The characteristic length for our parameters is $a_w=23.8$ nm for $B=0.1$ T.

The equilibrium properties of the closed system are found by diagonalizing its Hamitonian
(\ref{Hclosed}) stepwise in large bases. First, neglecting the cavity-photon interactions, by using 
a Fock space built from the noninteracting many-electron states of the system, $|\mu\rangle$, 
to obtain the spectrum and states, $|\mu )$, of the Coulomb interacting electrons.
We use the 36 lowest in energy single-electron states to build the noninteracting many-electron 
states with up to 3 electrons. The number of two- and three-electron states is selected  
according to their energy such that we have all states with energy up to 8 meV. The spin of
the electrons is included in this construction, and the energy threshold is set to cover 
states well above the bias window defined by the external leads.
Second, by building a Fock space as a tensor product of the lowest in energy 120 Coulomb
interacting states and the lowest 16 eigenstates of the photon number operator, the spectrum and properties of
the photon-dressed electron states, $|\breve{\mu})$, are calculated.\cite{Gudmundsson:2013.305}

\section{Equilibrium properties of the closed central system}
In order to study the transport of electrons through a two-dimensional nano-scale electronic system
with parallel quantum dots placed in a photon cavity in the regime of ultra strong electron-photon 
coupling we concentrate on two cases. The first case is the study of the transport through the 
one-electron ground state with the cavity-photon field close to a resonance with the 
ground state and the first excitation thereof. We will use the knowledge gathered from this 
case to study the more complex second case of the transport through the interacting 2-electron
ground state coupled to the first excitation thereof by a nearly resonant cavity photon field.  

Before engaging in this journey we need to explore special properties of the closed central system
brought around by the geometry of the system and the three interactions accounted for in the model.

\section{Resonance with the two lowest one-electron states}
\begin{figure}[htb]
      \includegraphics[width=0.45\textwidth]{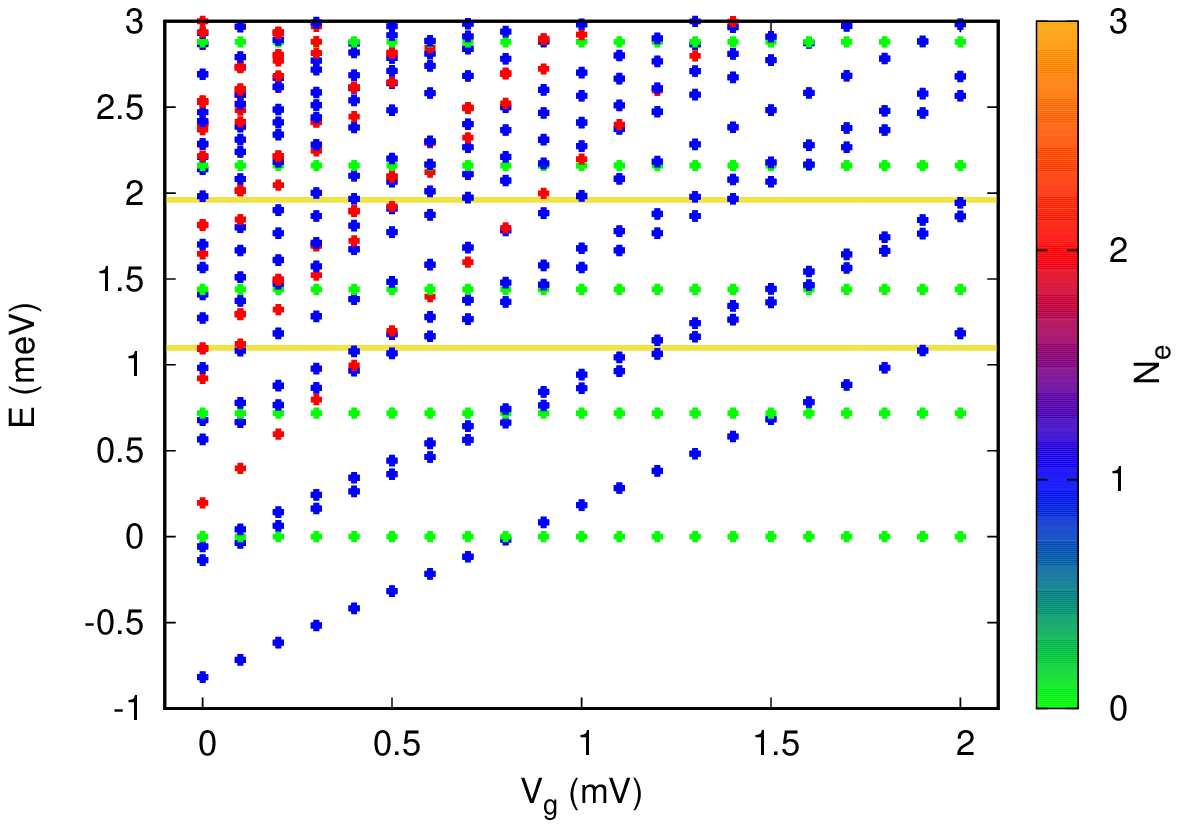}
      \includegraphics[width=0.45\textwidth]{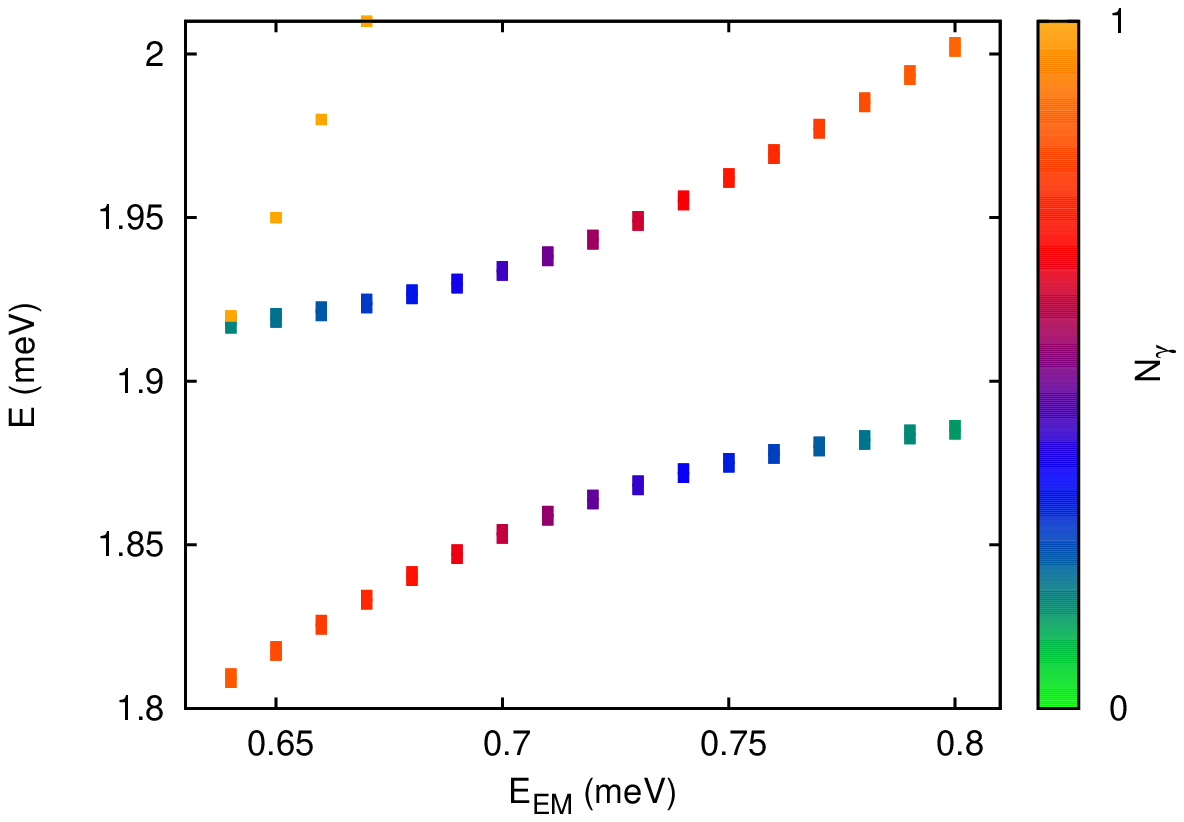}
      \caption{The many-body energy spectrum for the closed central system as function of 
               the plunger gate voltage $V_g$ for a $y$-polarized cavity
               photon field (upper panel), $\hbar\omega = 0.72$ meV. 
               The paramagnetic Rabi-splitting of the two spin components of the 
               first excitation of the one-electron ground state 
               as a function of the photon energy $E_\mathrm{EM}=\hbar\omega$ for $V_g=2.0$ mV (lower panel).
               $g_\mathrm{EM}=0.05$ meV. The horizontal yellow lines represent
               the chemical potentials of the left lead $\mu_L=1.10$ meV, and the right lead $\mu_L=1.96$ meV
               to be introduced below.}
      \label{Fig-Rabi-1e-y}
\end{figure}

The many-body energy spectrum for the central system as a function of the plunger gate voltage $V_g$ for the case of 
$y$-polarized cavity photons is displayed in the upper panel of Fig.\ \ref{Fig-Rabi-1e-y} with a color coding to 
indicate the electron content of each state. 
The eigenstates of the closed system can only contain an integer number of electrons. 
In our case it is in the range from 0 to 3.  
The fully interacting states of the central system, the cavity-photon dressed electron states,
are noted by $|\breve{\mu})$, with $\mu$ an integer (quantum number) assigned from 1 to
$N_\mathrm{mes}=120$ in a numerical ascending order determined by the energy of the state. 
As the upper panel of Fig.\ \ref{Fig-Rabi-1e-y} shows, the state number, $\mu$, thus 
depends on the plunger gate voltage $V_g$ and the photon energy $E_\mathrm{EM}=\hbar\omega$.    
The cavity-photon field has energy $\hbar\omega = 0.72$ meV, very close 
to the energy difference between both spin components of the one-electron ground state ($|\breve{3})$ 
and $|\breve{4})$) and the first excitation thereof.
A splitting of the first excitation of the one-electron state is seen in the upper panel of Fig.\ \ref{Fig-Rabi-1e-y}
(seen as states $|\breve{6})$ - $|\breve{9})$), 
and when viewed as a function of the photon energy $\hbar\omega$ in the lower panel of Fig.\ \ref{Fig-Rabi-1e-y} 
an anticrossing with an exchange of the photon content of the states appears indicating a typical Rabi-splitting. 
(The Zeeman spin splitting is not clearly visible, but can be deduced from the resulting symbols not being
square shape when overlapping with a slight off-set). 
The parameters selected for the parallel double quantum dots here lead to the one-electron wavefunction
of the first excited state to be an even function in the $x$-direction with no node, but uneven in the 
$y$-direction, with a nodal line situated between the dots. 
First order perturbation calculation shows that the paramagnetic electron-photon interaction, 
which also can be expressed as proportional to the integral of $\mathbf{r}\cdot\mathbf{E}$, where $\mathbf{E}$ is the 
electric field component of the cavity field, couples the one-electron ground state with its lowest lying excitation.
This is reminiscent of a dipole transition between the 1S and the 2P states of a Hydrogen atom promoted by a time varying
spatially constant electric field. 

\begin{figure}[htb]
      \includegraphics[width=0.45\textwidth]{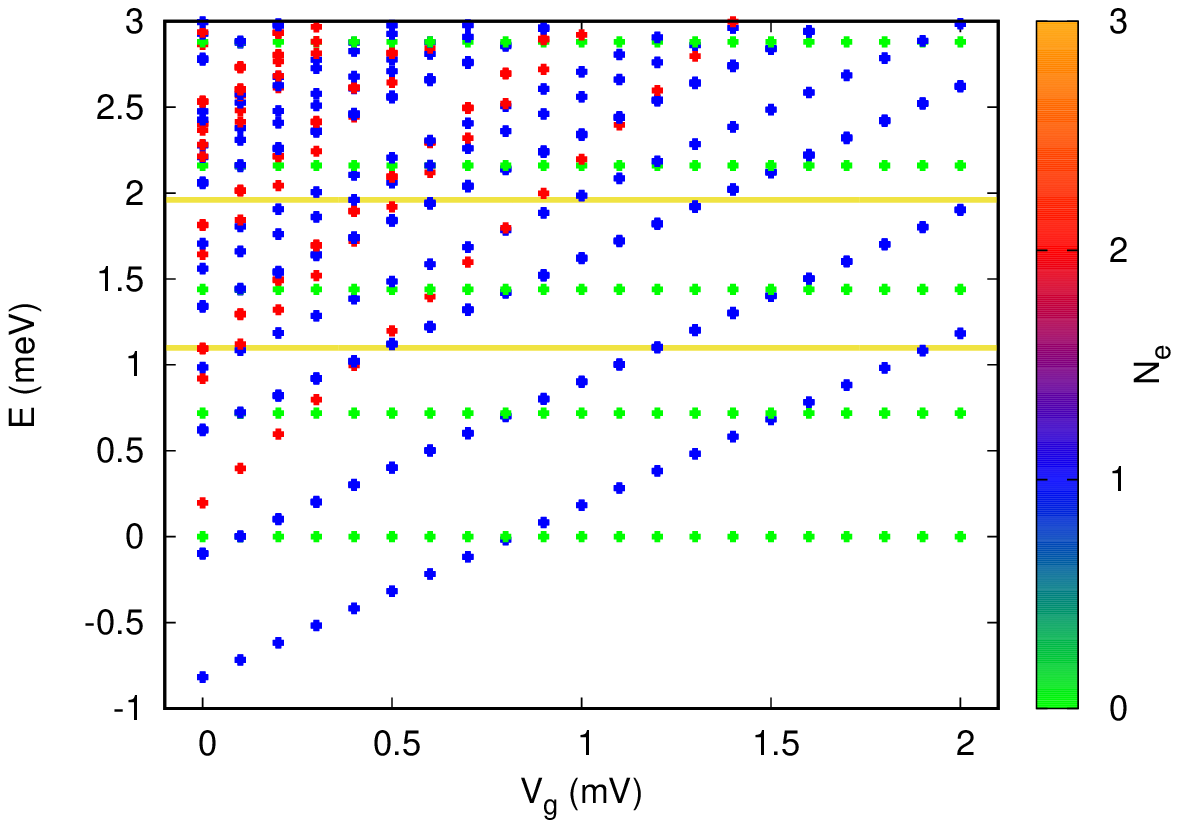}
      \includegraphics[width=0.45\textwidth]{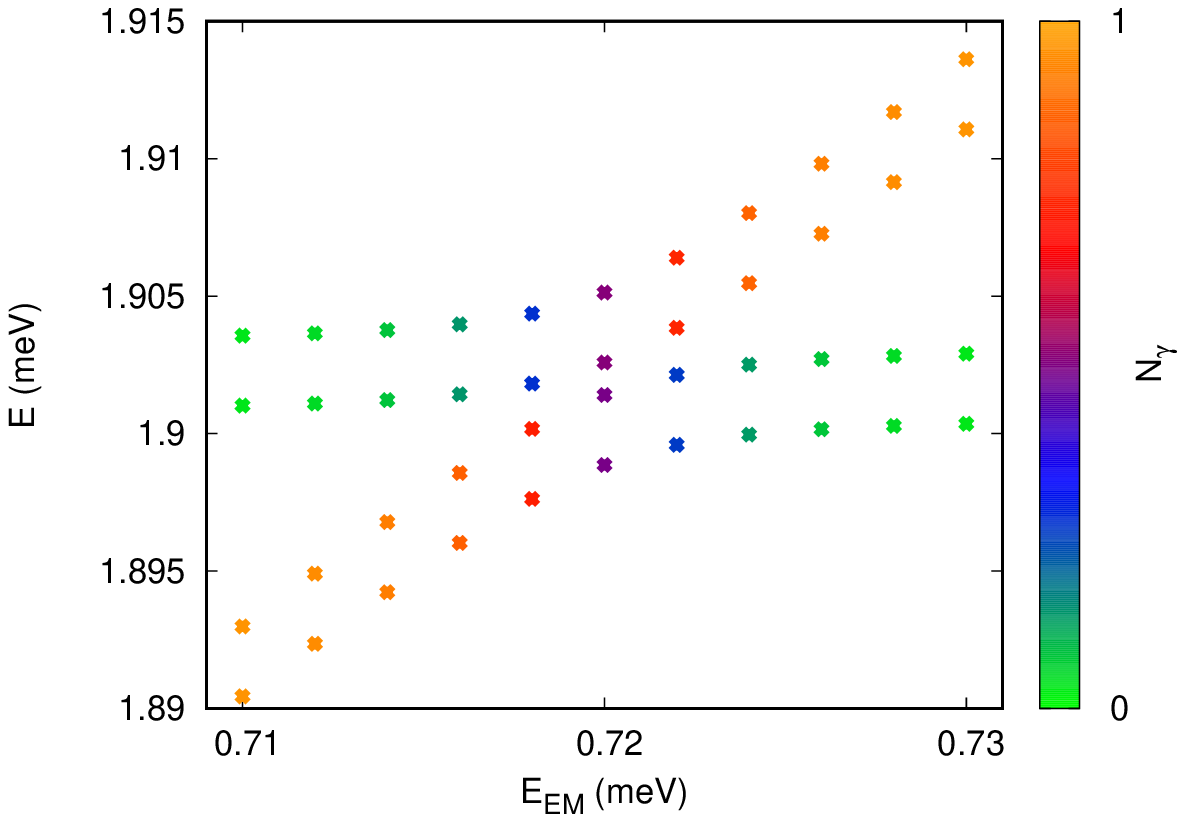}
      \caption{The many-body energy spectrum for the closed central system as function of 
               the plunger gate voltage $V_g$ for a $x$-polarized cavity
               photon field (upper panel), $\hbar\omega = 0.72$ meV.
               The diamagnetic Rabi-splitting of the two spin components of the 
               first excitation of the one-electron ground state 
               as a function of the photon energy $E_\mathrm{EM}=\hbar\omega$ for $V_g=2.0$ mV (lower panel).
               $g_\mathrm{EM}=0.05$ meV. The horizontal yellow lines represent
               the chemical potentials of the left lead $\mu_L=1.10$ meV, and the right lead $\mu_L=1.96$ meV
               to be introduced below.}
      \label{Fig-Rabi-1e-x}
\end{figure}

This analysis thus predicts that there is no Rabi-splitting expected for the same states 
(the two spin components of the one-electron ground state) for the case of an 
$x$-polarized cavity-photon field. States with the same parity along the $x$-axis can not be dipole coupled 
by the paramagnetic electron-photon interaction. 
The first impression of the upper panel of Fig.\ \ref{Fig-Rabi-1e-x} does not hint at
any splitting, but a closer analysis indicates a very weak splitting displayed in the lower panel
of Fig.\ \ref{Fig-Rabi-1e-x}. It is weak, as the splitting is of the same order of magnitude as 
the Zeeman spin splitting in GaAs at $B=0.1$ T. 
Due to the symmetry properties of the wavefunctions it is clear that this splitting is a Rabi-splitting
promoted by the diamagnetic electron-photon interaction, proportional to the integral of $\rho A^2$. 
This interaction is neglected in the original Jaynes-Cumming model,\cite{Jaynes63:89} but recently 
Malekakhlagh and T{\"u}reci show that in strongly coupled circuit-QED systems a corresponding term
should be considered.\cite{PhysRevA.93.012120} In atomic systems the weak effects of the diamagnetic 
interaction have been measured in the hyperfine diamagnetic shift of the ground state of 
$^9$Be$^+$.\cite{PhysRevA.84.012510}

\section{Resonance with the two lowest energy two-electron states}
\begin{figure}[htb]
      \includegraphics[width=0.45\textwidth]{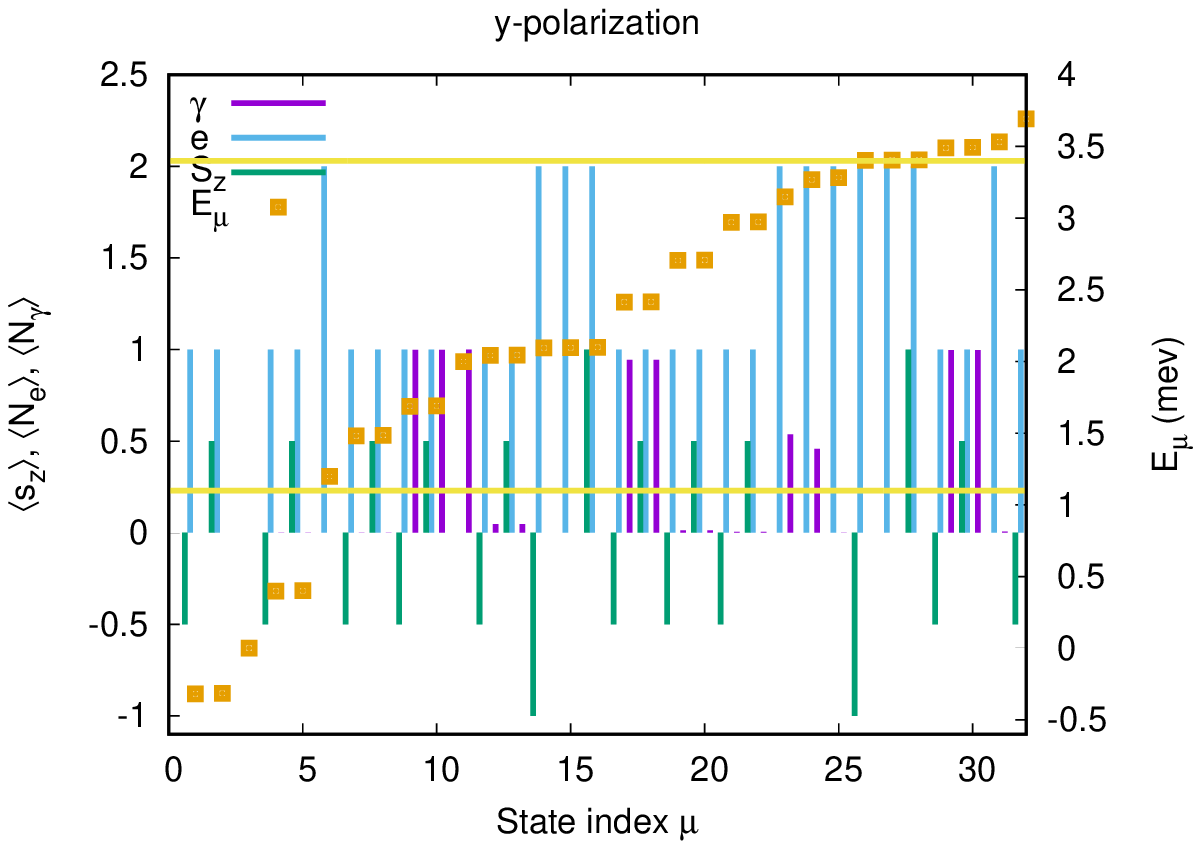}
      \includegraphics[width=0.45\textwidth]{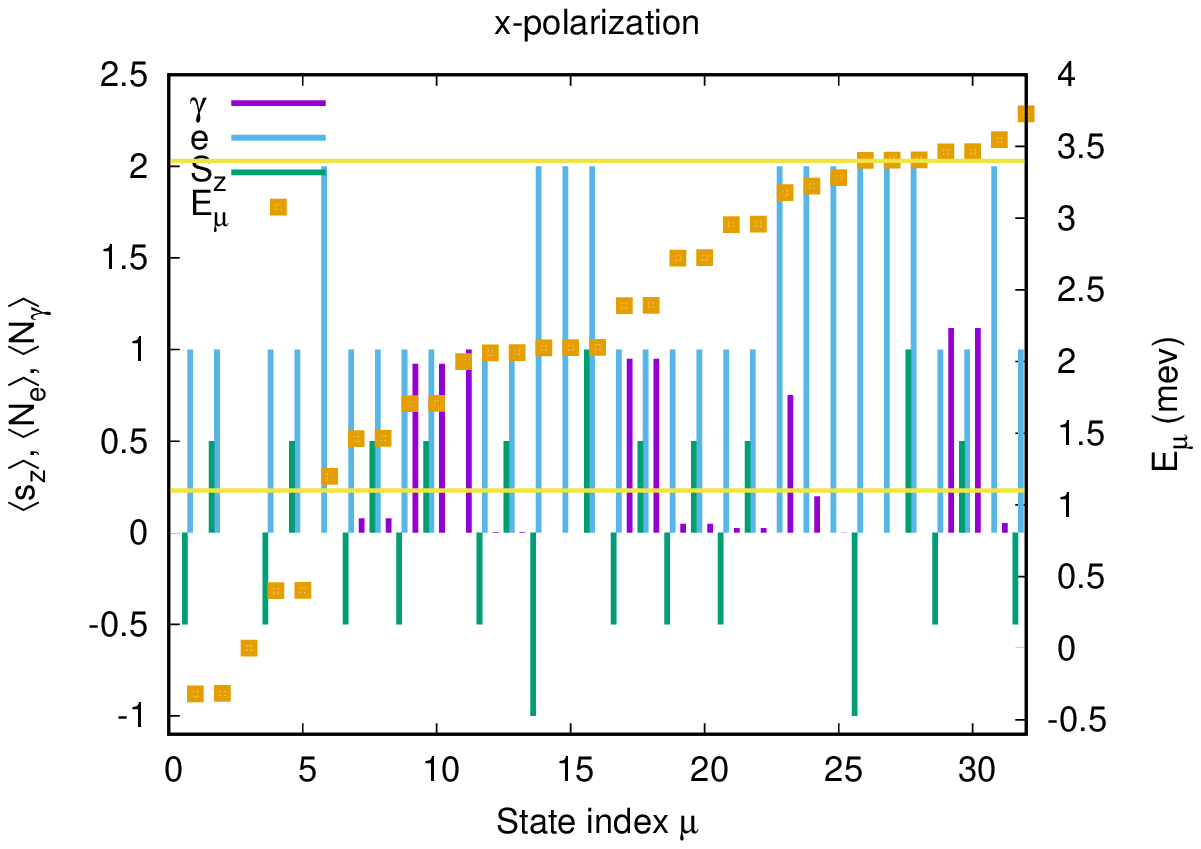}
      \caption{The properties of the many-body energy spectra for $V_g=0.5$ mV
               and $y$-polarized cavity photon field (upper panel), and 
               $x$-polarized cavity photon field (lower panel).
               The horizontal yellow lines represent
               the chemical potentials of the left lead $\mu_L=3.40$ meV, the right lead $\mu_R=1.40$ meV, 
               $\hbar\omega = 2.0$ meV, and $g_\mathrm{EM}=0.05$ meV.
               The squares indicate the energy $E_\mu$ of each state $|\breve{\mu})$, 
               and the impulses show the photon expectation value (labeled with $\gamma$), the electron number 
               (labeled with e), and the $z$-component of the spin ($S_z$).}
      \label{Fig-2e-structure}
\end{figure}

The structure of the lowest two-electron states of the system is more complex.
With cavity photons of energy $\hbar\omega=2.0$ meV there is a near resonance between
the photons and the two lowest spin-singlet two-electron states.
In Fig.\ \ref{Fig-2e-structure} we see the two-electron spin-singlet ground state, 
$|\breve{6})$, and the lowest lying triplet states, $|\breve{14})$, $|\breve{15})$, and $|\breve{16})$.
The noninteger photon content of the two-electron spin-singlet states $|\breve{23})$ and $|\breve{24})$
and their energy makes them strong candidates for the Rabi-split states resulting from the interaction
of the first photon replica of the two-electron ground state and the first excitation thereof. 
State $|\breve{25})$ is a two-electron singlet with higher charge distribution in the contact areas of
the short quantum wire, rather than in the quantum dots. States $|\breve{26})$, $|\breve{27})$, and 
$|\breve{28})$ are the corresponding triplet states. None of the last mentioned 4 states have appreciable
photon content. 

\begin{figure}[htb]
      \includegraphics[width=0.45\textwidth]{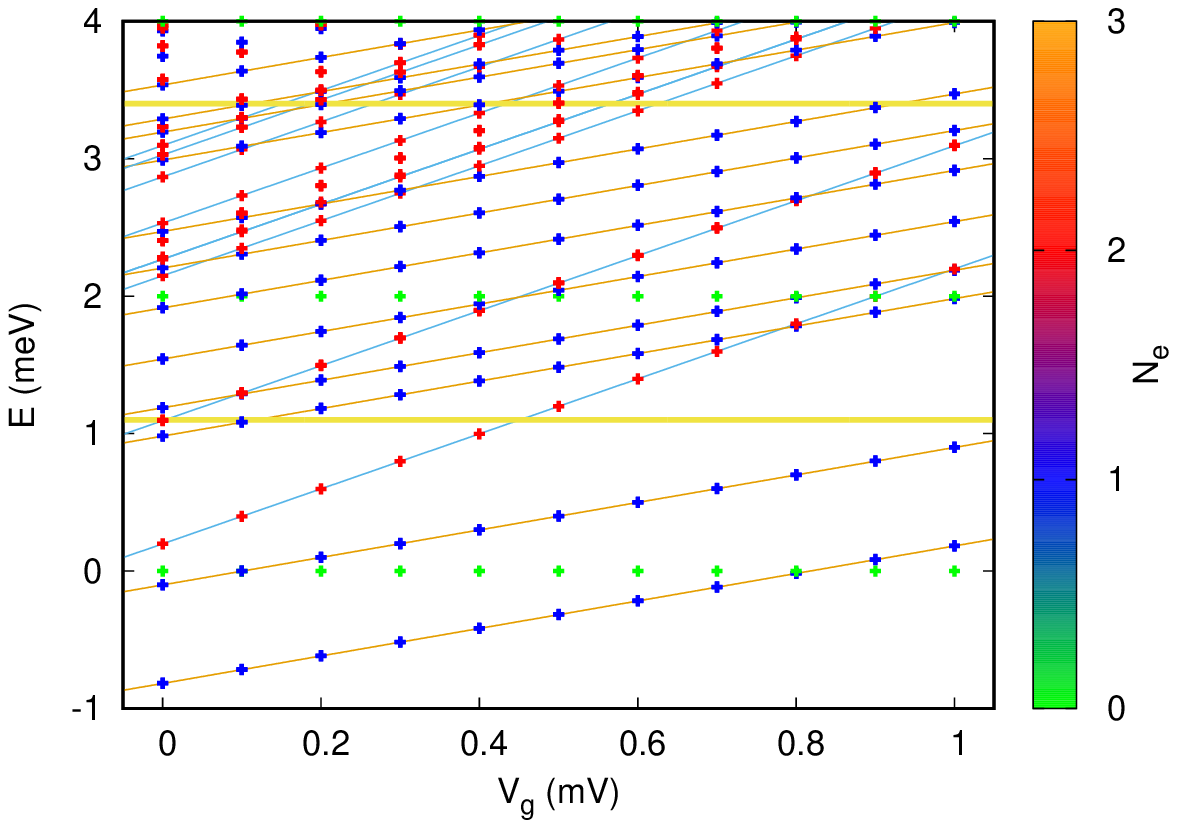}
      \includegraphics[width=0.45\textwidth]{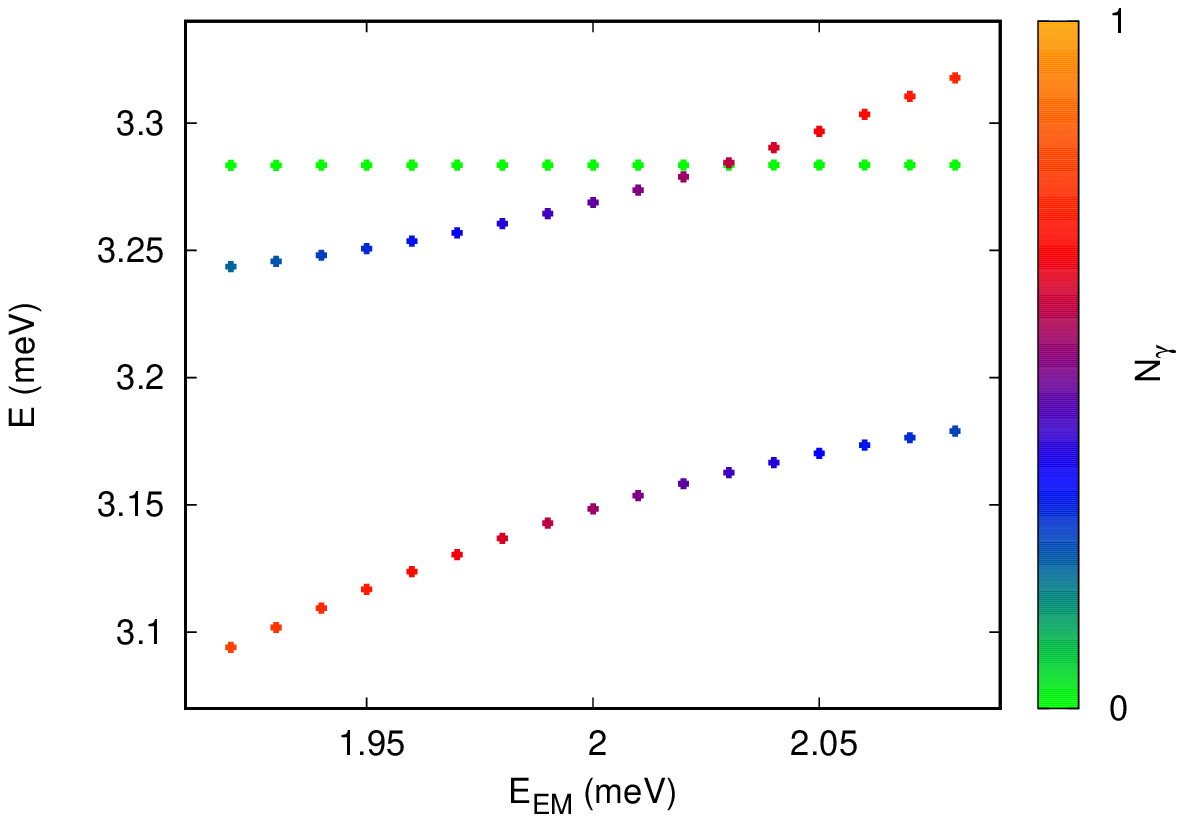}
      \caption{The many-body energy spectrum for the closed central system as function of 
               the plunger gate voltage $V_g$ for a $y$-polarized cavity
               photon field (upper panel), $E_\mathrm{EM}=\hbar\omega = 2.00$ meV. 
               The Rabi-splitting of the  
               first excitation of the two-electron ground state for $V_g=0.5$ mV (lower panel).
               $g_\mathrm{EM}=0.05$ meV. The horizontal yellow lines represent
               the chemical potentials of the left lead $\mu_L=3.40$ meV, and the right lead $\mu_R=1.40$ meV.}
      \label{Fig-Rabi-2e-y}
\end{figure}

Fig.\ \ref{Fig-Rabi-2e-y} presents in the upper panel the many-body spectrum for a $y$-polarized 
cavity-photon field as a function of the plunger gate voltage $V_g$, and in the lower panel 
the Rabi-splitting as a function of the photon energy of the states $|\breve{23})$ 
and $|\breve{24})$ for $V_g=0.5$ mV.

\begin{figure}[htb]
      \includegraphics[width=0.45\textwidth]{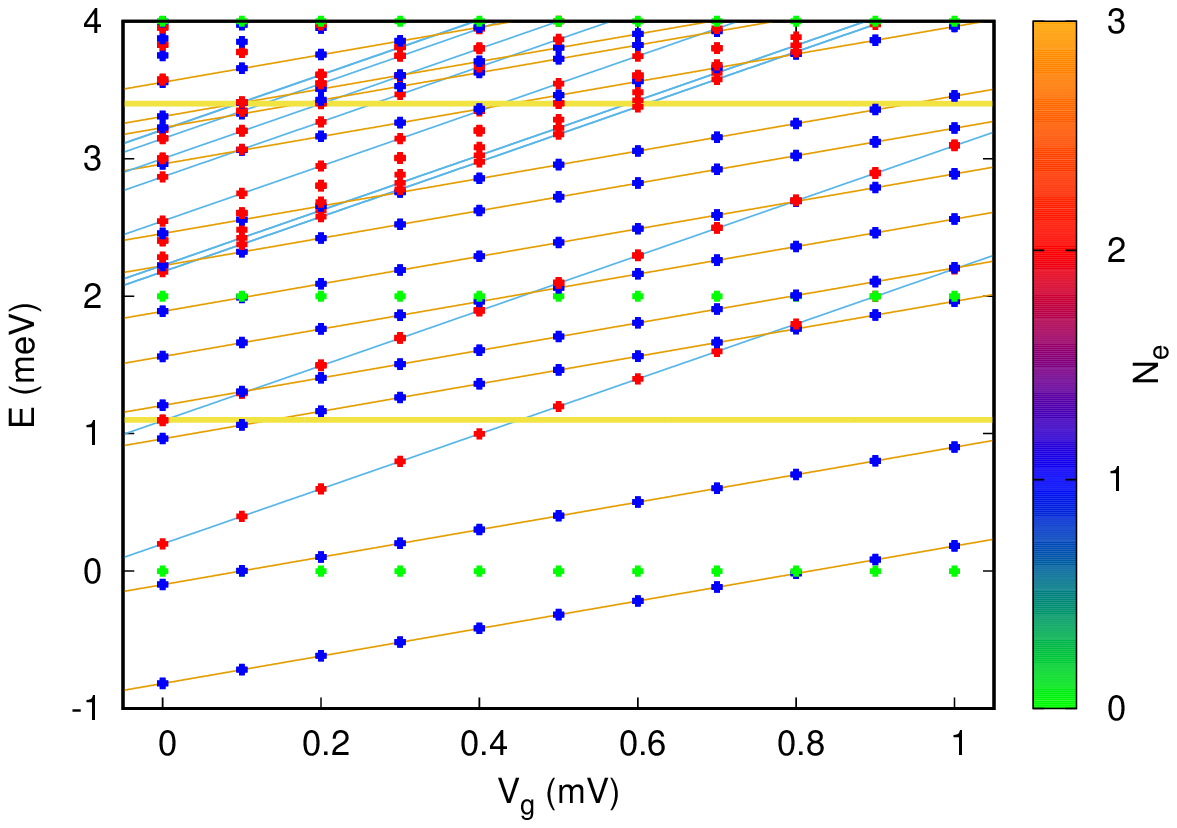}
      \includegraphics[width=0.45\textwidth]{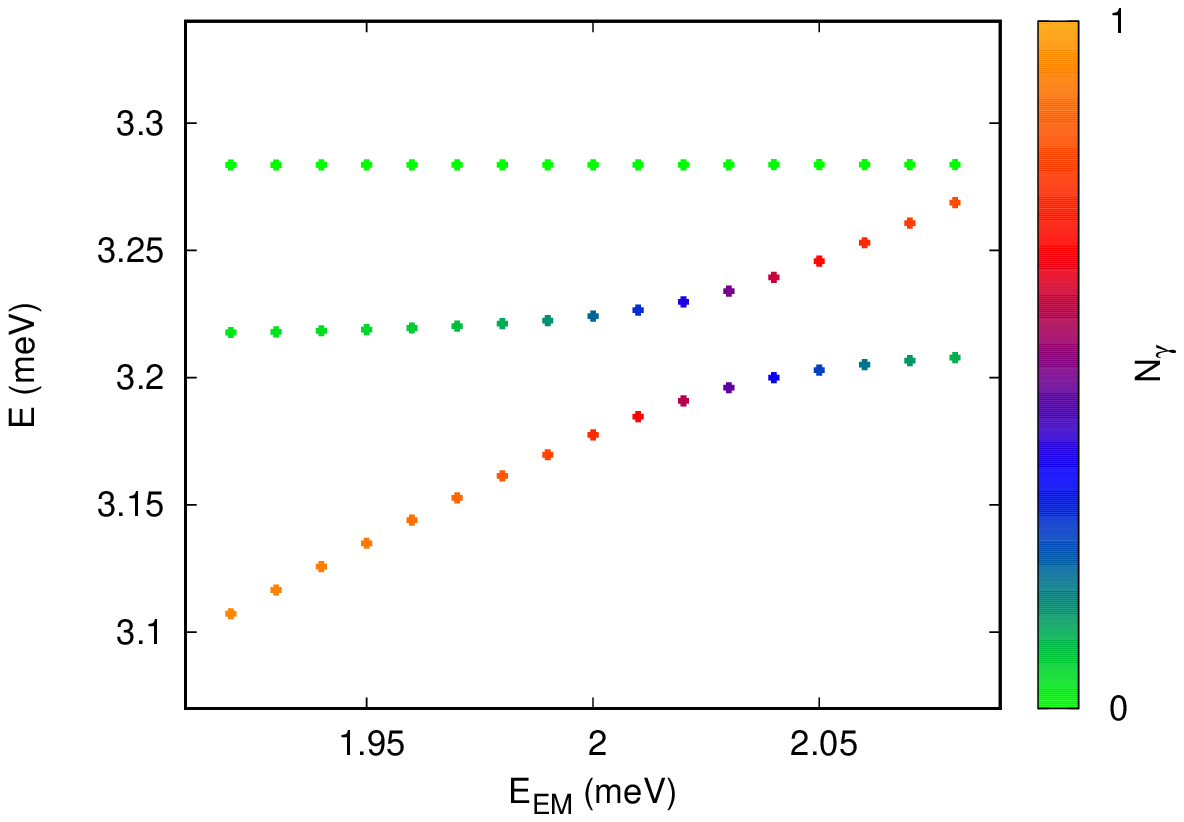}
      \caption{The many-body energy spectrum for the closed central system as function of 
               the plunger gate voltage $V_g$ for a $x$-polarized cavity
               photon field (upper panel), $\hbar\omega = 2.00$ meV. 
               The Rabi-splitting of the  
               first excitation of the two-electron ground state for $V_g=0.5$ mV (lower panel).
               $g_\mathrm{EM}=0.05$ meV. The horizontal yellow lines represent
               the chemical potentials of the left lead $\mu_L=1.10$ meV, and the right lead $\mu_L=3.40$ meV
               to be introduced below.
               In the upper panel the horizontal green dots represent purely photonic states
               with no electron component, but in the lower panel the state with horizontal 
               green dots has two electrons and no photon content.}
      \label{Fig-Rabi-2e-x}
\end{figure}

The results for the $x$-polarized cavity photons, but otherwise corresponding to the
case presented in Fig.\ \ref{Fig-Rabi-2e-y}, are displayed in Fig.\ \ref{Fig-Rabi-2e-x}.
Again, like for the case of one-electron states, we observe a larger Rabi-splitting for the $y$-polarization.
Symmetry arguments for the states again point to the diamagnetic electron-photon as the main cause for the 
Rabi-splitting for the case of $x$-polarized cavity photons. The two-electron states, nearly resonantly 
coupled by the cavity-photon field, are spin-singlet states and thus show no spin splitting, 
and higher electron density increases the strength of the diamagnetic electron-photon interaction compared 
to the case of one electron present in the system.

\section{Transport}
At the time point $t=0$ the central system is opened by coupling it to the left (L) and
the right (R) external leads acting like electron reservoirs with chemical potentials $\mu_L$
and $\mu_R$. The external semi-infinite leads are parabolically confined in the $y$-direction 
perpendicular to the transport and have a hard wall potential at the end. The leads have
the same characteristic confinement energy as the central system, $\hbar\Omega_0$, and are subjected
to the same external perpendicular magnetic field, B. The geometry and properties of the external
leads result in a continuous single-electron energy spectrum with a subband structure
characteristic for quasi-one dimensional electron systems.\cite{Gudmundsson:2013.305}   
The coupling to the leads depends on the geometry of the wavefunctions of the 
single-electron states in the {\lq\lq}contact area{\rq\rq} of the leads and the central system 
defined to extend approximately $a_w$ into each subsystem. In terms of creation and annihilation
operators of single-electron states in the leads ($c_{ql}^\dagger \mbox{ and } c_{ql}$) and the
central system ($d_{i}^\dagger \mbox{ and } d_{i}$) the coupling 
Hamiltonian is\cite{Gudmundsson09:113007,Moldoveanu10:155442,2016arXiv161003223J}
\begin{equation}
      H_\mathrm{T}(t)=\sum_{i,l}\int dq\;
      \left\{T^l_{qi}c_{ql}^\dagger d_i + (T^l_{qi})^*d_i^\dagger c_{ql} \right\},
\label{H_T}
\end{equation}
where the index $q$ stands for the combined continuous lead momentum quantum number and a
discrete subband index $n_l$, $i$ labels the single-electron states in the central system, and 
$T^l_{qi}$ is the state-dependent coupling tensor with $l=\{L,R\}$. The temperature of the 
electron reservoirs in the leads is $T=0.5$ K.

We use a formalism of Nakajima\cite{Nakajima58:948} and Zwanzig\cite{Zwanzig60:1338}
in which the dynamics of the whole system is projected on the central system leading to a generalized
master equation (GME)\cite{ANDP:ANDP201500298} 
\begin{equation}
      \partial_t{\rho_\mathrm{S}}(t) = -\frac{i}{\hbar}[H_\mathrm{S},\rho_\mathrm{S}(t)]
      -\frac{1}{\hbar}\int_0^t dt' K[t,t-t';\rho_\mathrm{S}(t')] 
\label{GME}
\end{equation}
for the reduced density operator of the central system describing statistical properties of the 
central system under influence of the external leads. The reduced density operator is defined by 
tracing out variables of the leads $\rho_\mathrm{S}(t)=\mathrm{Tr_{LR}}\{\rho_T (t)\}$.
Besides the external leads,
at $t=0$ the central system is as well coupled to a photon reservoir with the Markovian terms
\begin{align}
       &-\frac{\kappa}{2}(\bar{n}_R+1)\left\{2a\rho_sa^\dagger - a^\dagger a\rho_s - \rho_sa^\dagger a\right\}\nonumber\\
       &-\frac{\kappa}{2}(\bar{n}_R)\left\{2a^\dagger\rho_sa - aa^\dagger\rho_s - \rho_saa^\dagger\right\}
\end{align}
added to the master equation in the many-body Fock space (\ref{GME}), where 
$\kappa$ is the cavity photon decay constant. As we will investigate possible
vacuum radiation from the system, we set the mean number of photons of the reservoir $\bar{n}_R=0$
in order to avoid influx of photons from it. 

For the time evolution we use a Markovian master equation in Liouville space of 
transitions\cite{Weidlich71:325,Nakano2010,2016arXiv161003223J} derived from a non-Markovian Nakajima-Zwanzig 
equation,\cite{Nakajima58:948,Zwanzig60:1338} that includes the leads-central system coupling
(\ref{H_T}) up to second order in its integral kernel.\cite{Gudmundsson:2013.305,Gudmundsson09:113007}
As there are usually many available radiative transitions in the system, and not all in resonance
with the photon field, we do not use the rotating wave approximation for the electron-photon 
interaction in the central system. 

We use the Markovian master equation\cite{Gudmundsson16:AdP_10,2016arXiv161003223J} to obtain 
the long-time evolution and the steady state properties of the central system, weakly coupled 
to the leads. We investigate the transport through the one- and two-electron ground
states of the central system under two different conditions: (i) With the respective ground state
situated in a narrow bias window, much smaller than the photon energy; (ii) with the ground state
situated in a bias window that is large enough to include its first photon replica, i.e.\ the photon
energy is smaller than the bias window. In the former case, energy of an incoming electron is
not sufficient to generate a cavity photon via standard electroluminescence, but in the latter one 
the energy is sufficient for that process.\cite{PhysRevLett.116.113601} 

We use the Fourier spectrum of the emitted cavity radiation of the
system in its steady state to build its spectral density as
\begin{equation}
      S(E) = \frac{\kappa}{\pi}\left|\int_0^\infty \frac{d\tau}{\hbar}e^{-iE\tau /\hbar}
             \{\langle X (\tau) X(0)\rangle\}\right| , 
\label{SE}
\end{equation}   
where $X=a+a^\dagger$, and the time point $\tau =0$ is now considered to mark any time after the onset 
of the steady state.\cite{PhysRevA.74.033811,PhysRevA.80.053810,PhysRevLett.116.113601}
According to the quantum regression theorem valid for a Markovian system weakly coupled to a reservoir 
the equation of motion for a two-time correlation function, as is in the
integrand of Eq.(\ref{SE}), is of the same type as the master equation for the reduced
density operator of the system, but for an effective density 
operator.\cite{Scully97,Carmichael99,Wallis-QO} Here, the effective density operator
is\cite{doi:10.1063/1.3570581}
\begin{equation}
      \chi (\tau ) = \mathrm{Tr}_\mathrm{res}\left\{ e^{-iH\tau /\hbar}X\rho_T(0)e^{+iH\tau /\hbar}\right\} ,
\end{equation}
with $H$ the Hamiltonian of the total system, $\rho_T$ its density operator, and $\mathrm{Tr}_\mathrm{res}$ the
trace operator with respect to the variables of the reservoir. In the Liouville space the solution 
of the Master equation is
\begin{equation}
      \mathrm{vec}({\chi (\tau )}) = \left\{ {\cal U}\left[\exp\left(
                         {\cal L}_\mathrm{diag}\tau \right) \right] {\cal V} \right\}
                         \mathrm{vec}(\chi (0))
\end{equation}
and the two-time average or the correlation function becomes
\begin{equation}
      \langle X (\tau ) X(0)\rangle = \mathrm{Tr}_\mathrm{S}\left\{ X(0) \chi (\tau ) \right\} . 
\end{equation}
Here, ${\cal L}$ is an approximation of the Liouville operator of the total system,  
${\cal L}_{\mathrm{diag}}$ is the complex diagonal matrix corresponding to it in the 
Liouville space of transitions, and ${\cal U}$ is the matrix of its left eigenvectors,
and ${\cal V}$ the matrix of its right eigenvectors.\cite{2016arXiv161003223J,PhysRevB.81.155303}
$\mathrm{Tr}_\mathrm{S}$ is the trace operation with respect to the state space of the central system.

\subsection{Electroluminescence due to transport through the one-electron ground state}
\begin{figure}[htb]
      \includegraphics[width=0.45\textwidth]{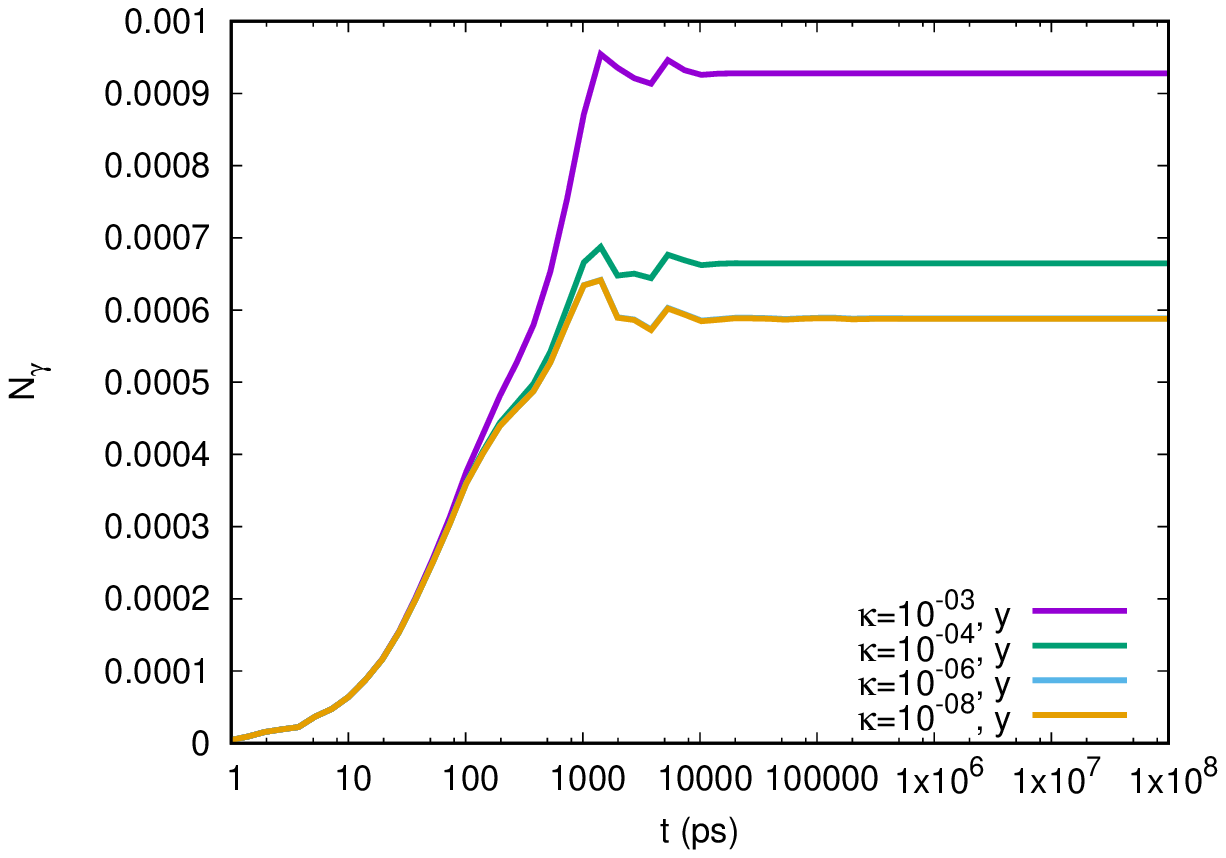}
      \includegraphics[width=0.45\textwidth]{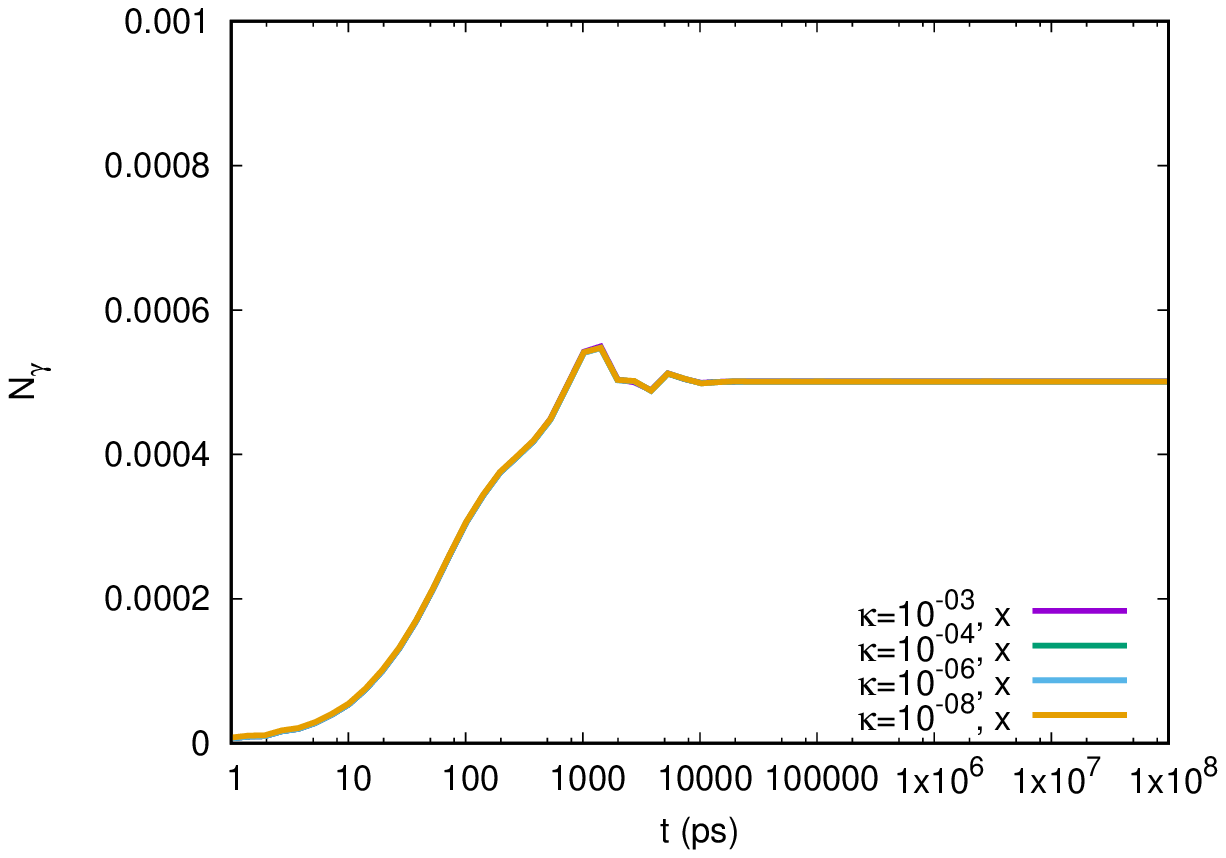}
      \caption{The mean photon number $N_\gamma$ as a function of time for the central system 
               initially in its vacuum state $|\breve{3})$, and $y$-polarized (upper panel), 
               or $x$-polarized (lower panel) cavity-photon field. 
               In the upper panel the curves for the two lowest values of $\kappa$
               overlap, and in the lower panel all curves overlap.
               $g_\mathrm{EM}=0.05$ meV, $V_g=2.0$ meV, $\hbar\omega =0.72$ meV,
               $\mu_R=1.1$ meV, and $\mu_L=1.4$ meV. The coupling to the photon reservoir,
               $\kappa$ is measured in meV.}
      \label{Ng-1e-muL1p40}
\end{figure}

In Fig.\ \ref{Ng-1e-muL1p40} the mean value of the photon number operator, 
$N_\gamma = \langle a^\dagger a \rangle$, is
shown as a function of time for the initially empty system (in state $|\breve{3})$) in case
of a small bias window $\Delta\mu =0.3$ meV ($\mu_L=1.4$ meV and $\mu_R=1.1$ meV) 
and plunger gate voltage $V_g=2.0$ mV. Here, only 
the two spin components of the one-electron ground state are inside the bias window and 
the photon energy $\hbar\omega =0.72$ meV is larger than the bias window $\Delta\mu$.
The initial rise in $N_\gamma$ for $t<10$ ns reflects the charging of the weakly coupled system, 
and the value for later time is, to the largest extent, determined by the steady-state charge in the 
system and the tiny photon component present in the one-electron ground state due to the 
electron-photon coupling, especially for the case of $x$-polarized photon field seen
in the lower panel of Fig.\ \ref{Ng-1e-muL1p40}, where $N_\gamma$ is independent of the
photon-cavity decay rate $\kappa$. 
Clearly, the $y$-polarized photon field couples 
stronger to the one electron in the central system as is seen in the upper panel of 
Fig.\ \ref{Ng-1e-muL1p40}. That is in accordance with the normalized coupling constant $\eta$
being larger as the respective Rabi-splitting seen in Fig.\ \ref{Fig-Rabi-1e-y} is larger in
this case. 
Moreover, a slight enhancement of $N_\gamma$ is visible for larger values of the photon-cavity 
decay rate $\kappa$. A close inspection of the state occupancy in the steady state, surprisingly, 
reveals that for the larger $\kappa$ the first photon replicas of the one-electron ground state 
gain a slight occupation for the $y$-polarized photon field, but not for the $x$-polarization.
In addition, the one-photon state, $|\breve{3})$ also acquires a slight occupation.
These states have a higher photon number resulting in a higher overall $N_\gamma$. 
This effect has to be termed a purely {\lq\lq}dynamic{\rq\rq} effect as it is not easily predictable from the
properties of the isolated central system. 

\begin{figure}[htb]
      \includegraphics[width=0.45\textwidth]{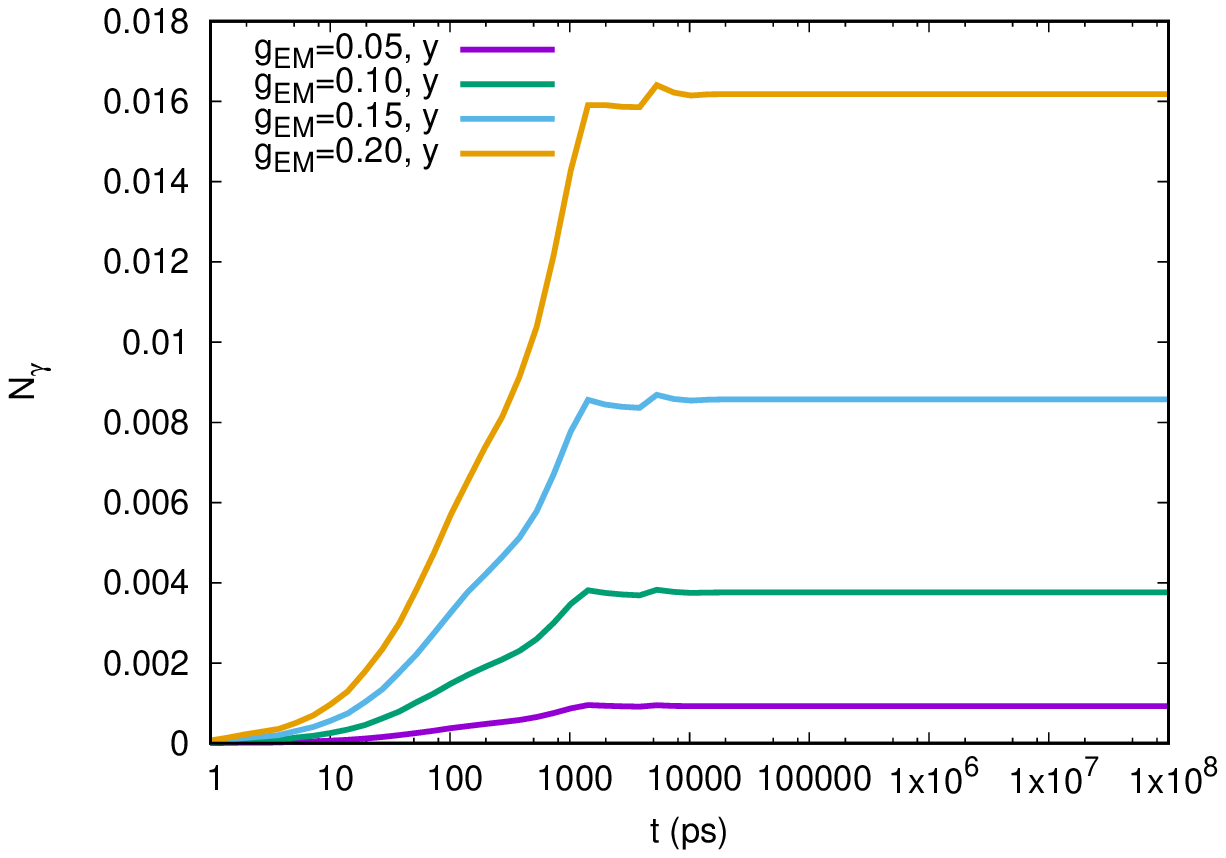}
      \includegraphics[width=0.45\textwidth]{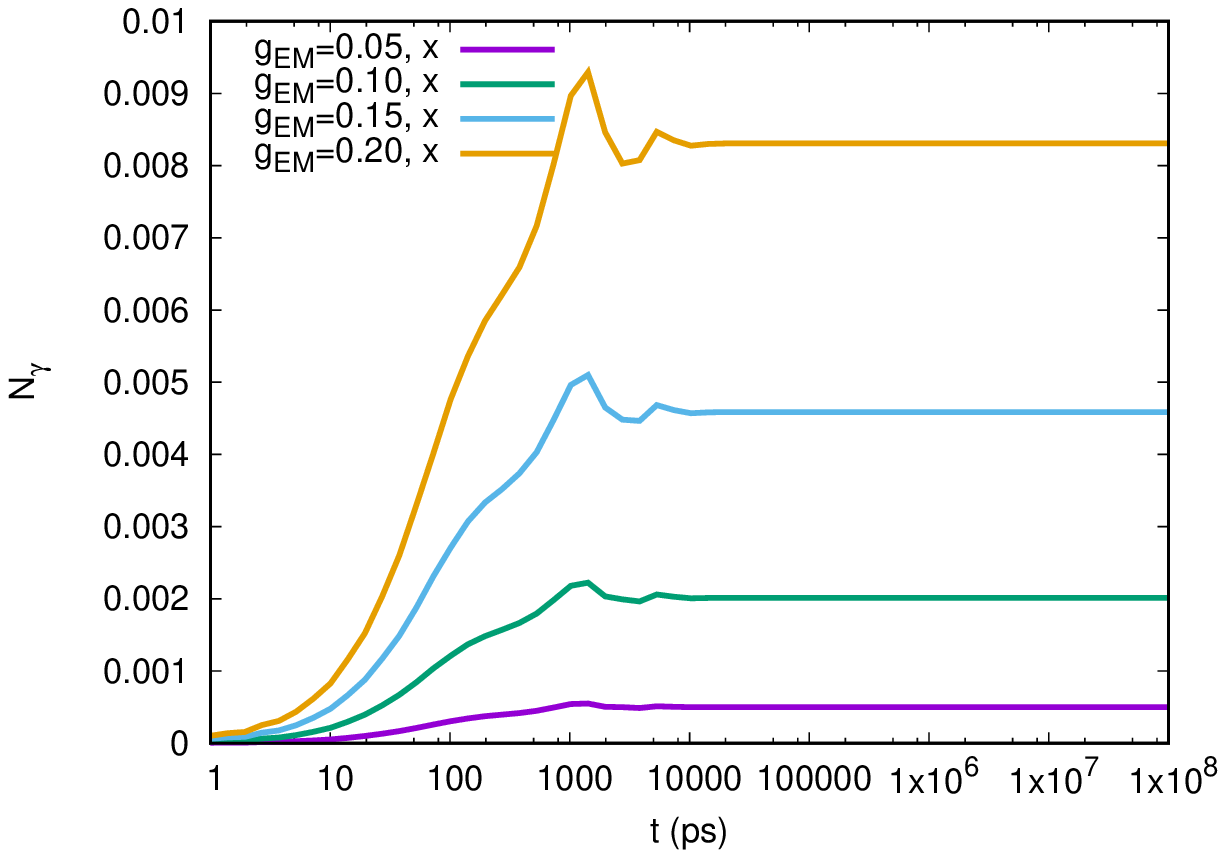}
      \caption{The mean photon number $N_\gamma$ as a function of time and the electron-photon 
               coupling strength $g_\mathrm{EM}$ measured in meV for the central system 
               initially in its vacuum state $|\breve{3})$, and $y$-polarized (upper panel), 
               or $x$-polarized (lower panel) cavity-photon field. 
               $\kappa=10^{-3}$ meV, $V_g=2.0$ meV, $\hbar\omega =0.72$ meV,
               $\mu_R=1.1$ meV, and $\mu_L=1.4$ meV.}
      \label{Ng-1e-muL1p40-varg}
\end{figure}

The nonlinear increase of the mean photon number $N_\gamma$ is displayed in 
Fig.\ \ref{Ng-1e-muL1p40-varg}, where again the stronger effective coupling 
is evident for the $y$-polarized photon field. Note again that here only the two spin components
of the one-electron ground state are inside the narrow bias window.

\begin{figure}[htb]
      \includegraphics[width=0.45\textwidth]{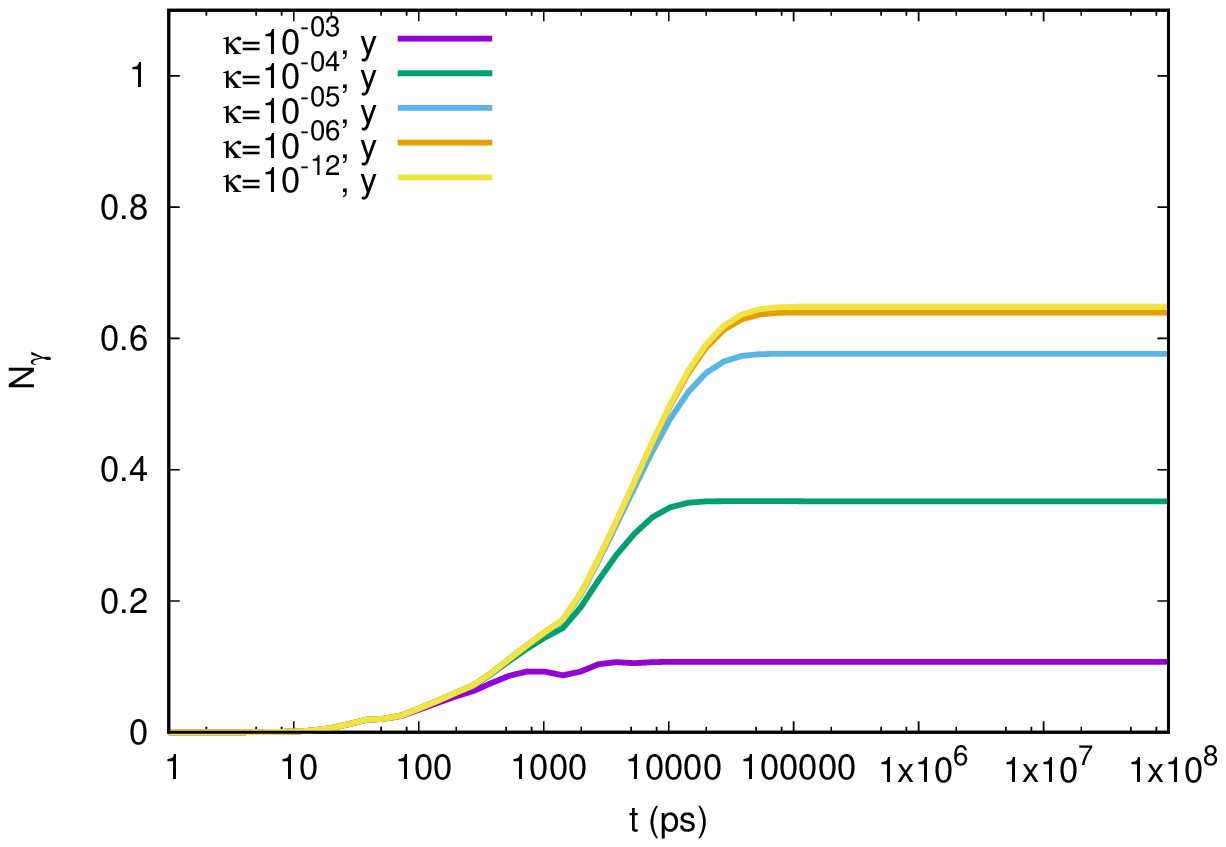}
      \includegraphics[width=0.45\textwidth]{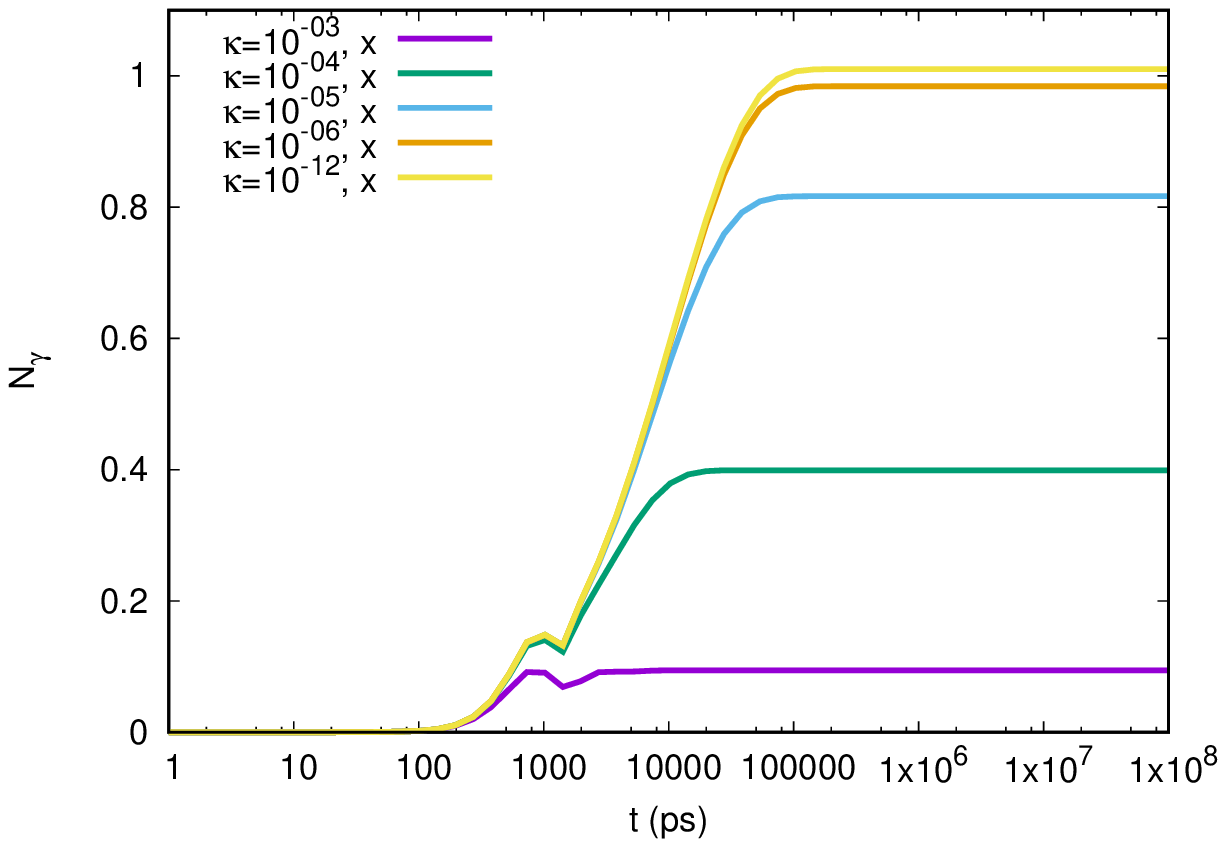}
      \caption{The mean photon number $N_\gamma$ as a function of time for the central system 
               initially in its vacuum state $|\breve{1})$, and $y$-polarized (upper panel), 
               or $x$-polarized (lower panel) cavity-photon field. 
               $g_\mathrm{EM}=0.05$ meV, $V_g=2.0$ meV, $\hbar\omega =0.72$ meV,
               $\mu_R=1.10$ meV, and $\mu_L=1.96$ meV. The coupling to the photon reservoir,
               $\kappa$ is measured in meV.}
      \label{Ng-1e-muL1p96}
\end{figure}

Extension of the chemical potential of the left lead, $\mu_L$, to just above the first photon
replica of the one-electron ground state to the value of 1.96 meV radically changes the situation
as is displayed in Fig.\ \ref{Ng-1e-muL1p96}.
Now the mean photon value, $N_\gamma$, assumes considerable values for both the 
$y$-polarized cavity photon (upper panel of Fig.\ \ref{Ng-1e-muL1p96}) and $x$-polarization
thereof (the lower panel). As could be expected, $N_\gamma$ increases with lower cavity-decay
rate $\kappa$ due to their accumulation in the central system. 

\begin{figure}[htb]
      \includegraphics[width=0.45\textwidth]{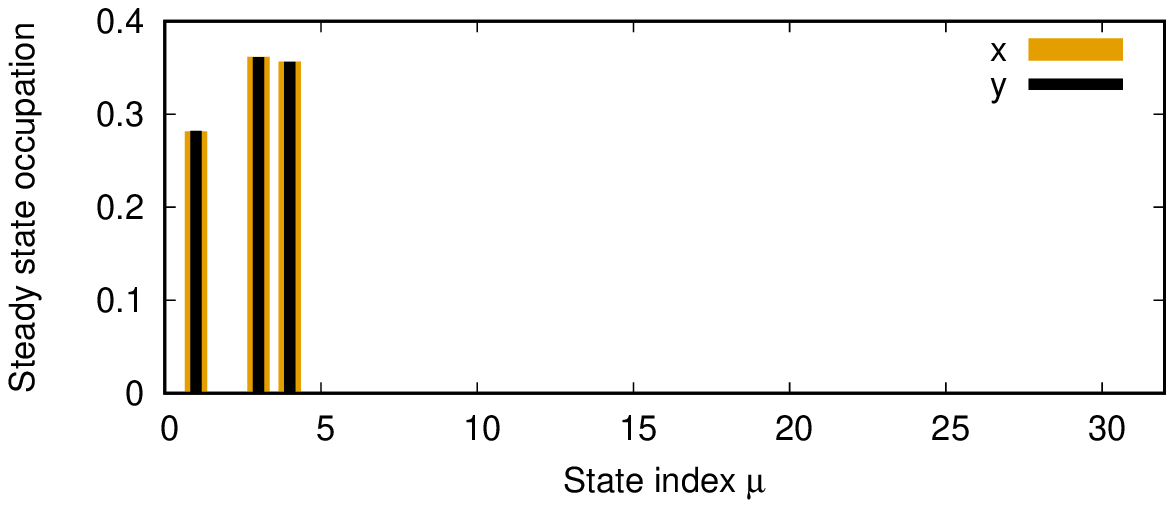}
      \includegraphics[width=0.45\textwidth]{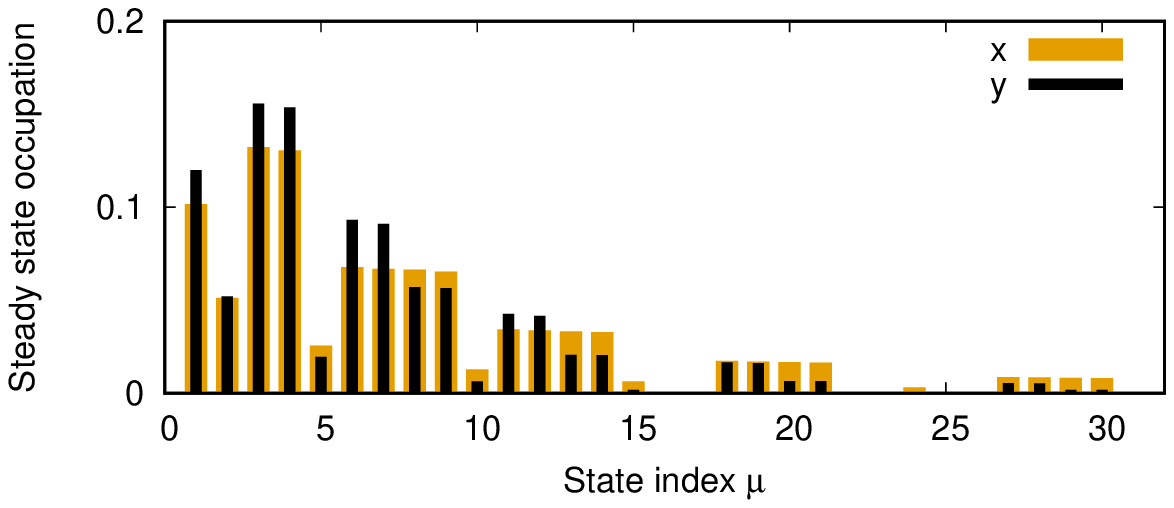}
      \caption{The steady-state occupation of the lowest 32 many-body states $|\breve{\mu})$,
               for the $x$- and $y$-polarized cavity photon field, for
               $V_g=2.0$ mV, $\mu_R=1.10$ meV, and $\mu_L=1.40$ meV (upper panel), and 
               $\mu_L=1.96$ meV (lower panel). The bias window includes states 
               $|\breve{3})$ - $|\breve{9})$. $g_\mathrm{EM}=0.05$ meV, 
               and $\kappa = 1\times 10^{-6}$ meV.}
      \label{SS-occ-1e}
\end{figure}

The steady-state occupation of the states of the central system is shown in Fig.\ \ref{SS-occ-1e}
for both polarizations, the smaller bias window with $\Delta\mu =0.3$ (upper panel), 
and the larger bias with $\Delta\mu =0.86$ meV (lower panel). 
The upper panel of Fig.\ \ref{SS-occ-1e} indicates that only the vacuum, $|\breve{1})$, and
both spin components of the one-electron ground state, $|\breve{2})$ and $|\breve{3})$,
acquire considerable occupation, whereas, the lower panel shows considerable occupation of states
above the bias window (states with a number higher than 9). 

The lower panel of Fig.\ \ref{SS-occ-1e} helps to understand the dependence of the photon
accumulation on the polarization seen in Fig.\ \ref{Ng-1e-muL1p96}. The Rabi-split states
for the $y$-polarization (the first photon replicas of the two spin states of the one-electron ground state),
$|\breve{6})$, $|\breve{7})$, $|\breve{8})$, and  $|\breve{9})$ are all just
under $\mu_L$ for the $x$-polarization, but two thereof, $|\breve{8})$, and  $|\breve{9})$,
touch $\mu_L$ for the case of $y$-polarization. For the $y$-polarization we thus see a slightly higher
occupation of states with lower photon component, $|\breve{1})$, $|\breve{3})$, $|\breve{4})$,
$|\breve{6})$, and  $|\breve{7})$. Notice furthermore, that in the lower panel of 
Fig.\ \ref{SS-occ-1e} the pure photon states with no electron component, $|\breve{2})$,
$|\breve{5})$, $|\breve{10})$, and $|\breve{15})$ all have a slight occupancy 
visible, but not in the upper panel. The steady-state occupation for the narrow bias window 
seen in the upper panel of Fig.\ \ref{SS-occ-1e} shows no visible difference between the two
polarization directions of the photon field, but the difference for the broad bias window, seen
in the lower panel, is caused by the interplay of the geometry and the electron-photon interactions. 
In the $x$-polarization the selection rules favor the diamagnetic part of the interaction, 
while in the $y$-polarization the paramagnetic interaction dominates. 

\begin{figure}[htb]
      \includegraphics[width=0.45\textwidth]{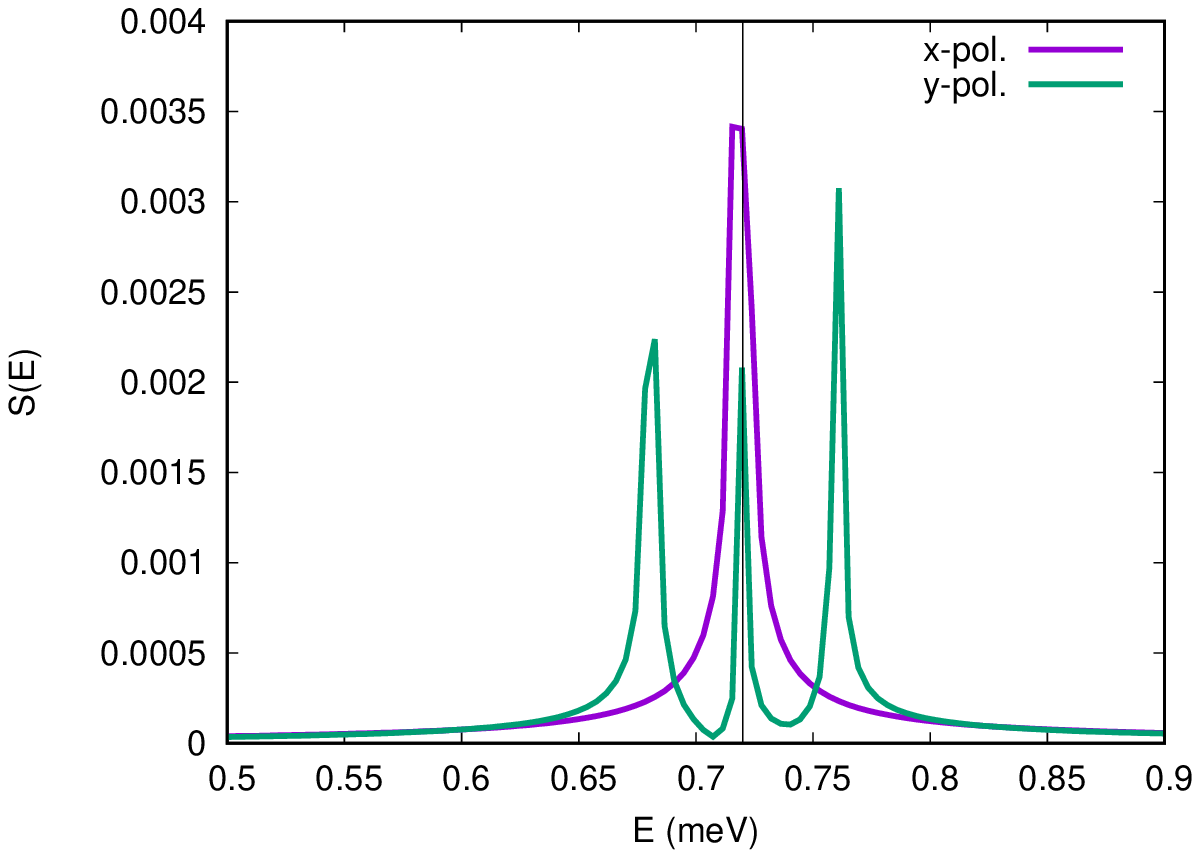}
      \includegraphics[width=0.45\textwidth]{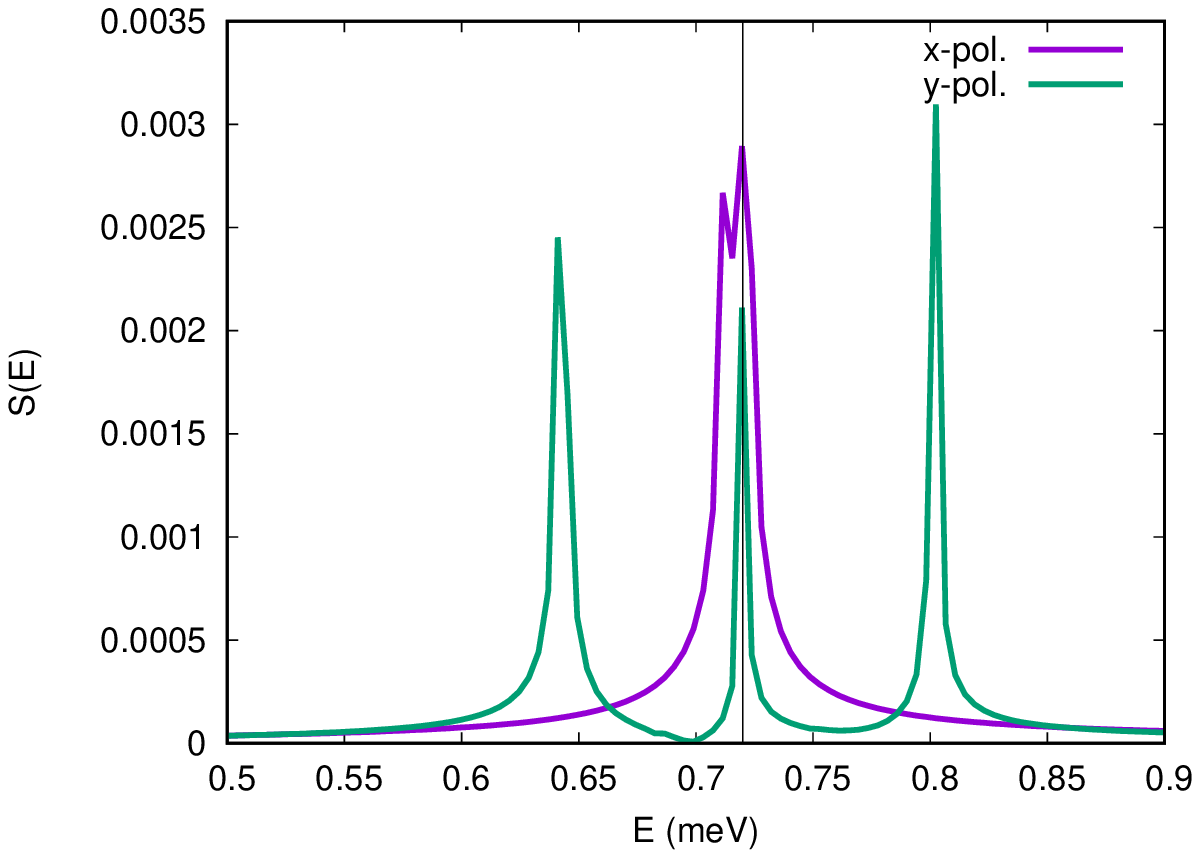}
      \caption{The spectral density $S(E)$ of the emitted cavity radiation of the system in its steady
               state for the electron-photon coupling $g_\mathrm{EM}=0.05$ meV (upper panel),
               and $g_\mathrm{EM}=0.10$ meV (lower panel). 
               $V_g=2.0$ mV, $\hbar\omega =0.72$ meV (vertical black line), $\mu_L=1.4$ meV, $\mu_R=1.1$ meV, 
               and $\kappa = 1\times 10^{-3}$ meV.}
      \label{SS-fft-1G}
\end{figure}

\begin{figure}[htb]
      \includegraphics[width=0.45\textwidth]{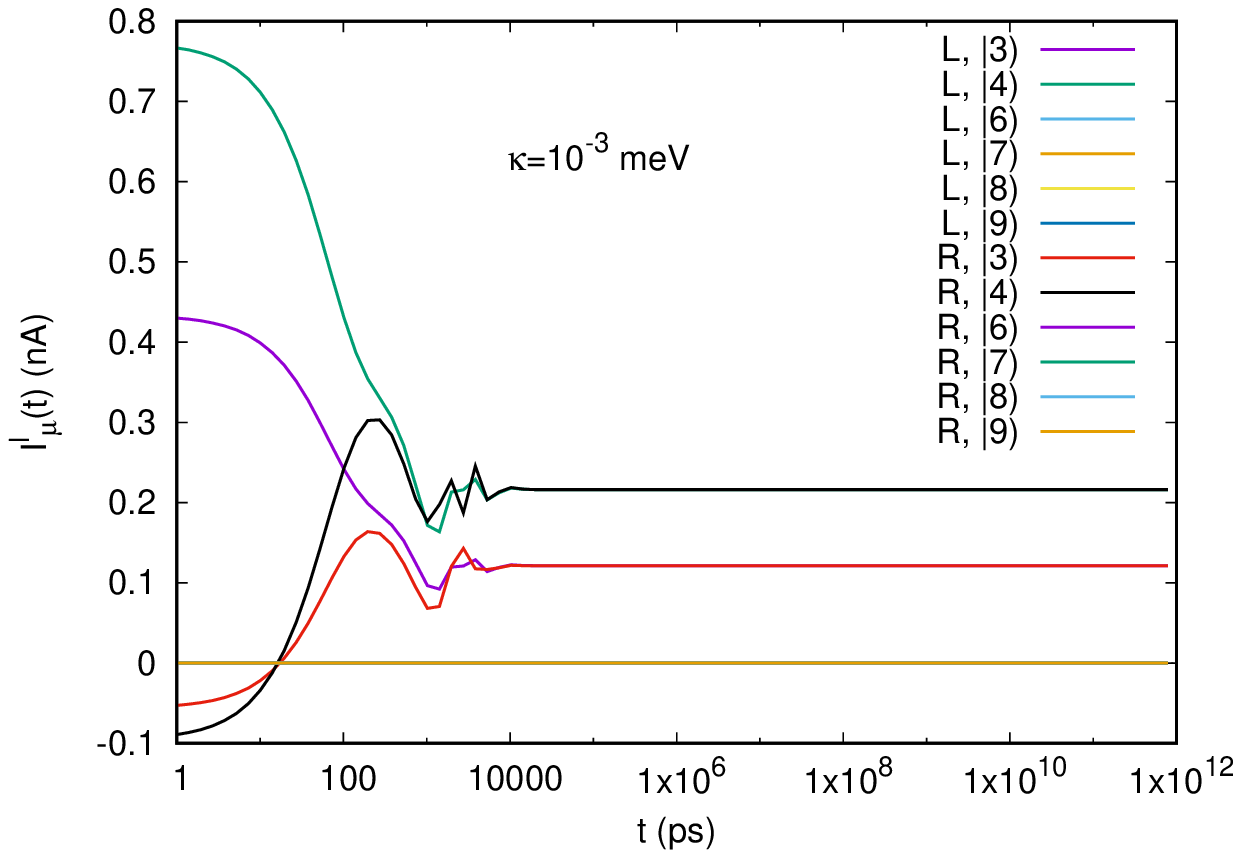}
      \includegraphics[width=0.45\textwidth]{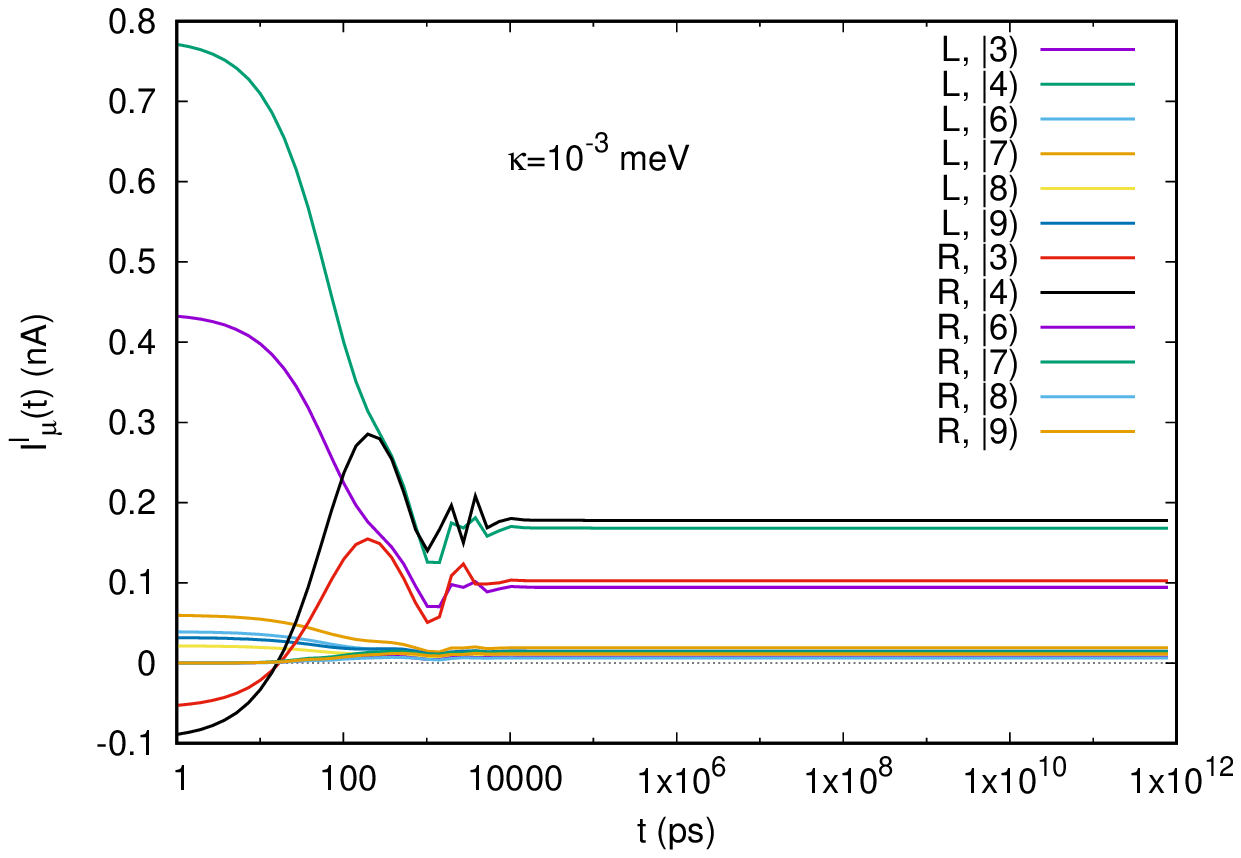}
      \caption{The partial current from the left lead (L) into state $|\breve{\mu})$, and the
               partial current from state $|\breve{\mu})$ into the right lead (R) as function
               of time for the narrow bias window, $\Delta\mu = 0.3$ meV, (upper panel), 
               and the broad bias window, $\Delta\mu = 0.86$ meV, (lower)
               panel. $V_g=2.0$ mV, $\hbar\omega =0.72$ meV, $g_\mathrm{EM}=0.05$ meV, 
               and $\kappa = 1\times 10^{-3}$ meV. $y$-polarized cavity photon field.} 
      \label{SS-JLR-1G}
\end{figure}

In order to analyze further the mixture of the standard polaritonic and the ground state 
electroluminescence present in the system for the plunger gate voltage $V_g=2.0$ mV 
and the photon energy $\hbar\omega =0.72$ meV, we present in Fig.\ \ref{SS-fft-1G}
the Fourier spectrum of the emitted cavity radiation of the system in its steady
state, and in Fig.\ \ref{SS-JLR-1G} the partial current through the relevant transport 
states.

The upper panel of Fig.\ \ref{SS-fft-1G} reveals for the $y$-polarized cavity photon, already at 
$g_\mathrm{EM}=0.05$ meV, the well known Mollow triplet,\cite{PhysRev.188.1969,PhysRevB.85.115309} 
with the central peak representing the emission of the cavity at the fundamental mode, 
$\hbar\omega =0.72$ meV, and the two satellites representing emission at the two Rabi-shifted frequencies. 
For the $x$-polarization of the 
cavity field, only a hint of splitting is seen for $g_\mathrm{EM}=0.05$ meV in the upper panel of 
Fig.\ref{SS-fft-1G}, but it becomes clearer at $g_\mathrm{EM}=0.1$ meV seen in the lower panel.
A fully developed two-peak structure is found at $g_\mathrm{EM}=0.2$ meV (not displayed here) 
with the higher energy peak located at the fundamental energy $\hbar\omega =0.72$ meV.
This is in accordance with the small Rabi splitting for the $x$-polarized cavity field
seen in Fig.\ \ref{Fig-Rabi-1e-x} and the fact that it is produced by the diamagnetic 
electron-photon interaction.

Fig.\ \ref{SS-JLR-1G} for the partial currents into and from the central system for the 
case of a $y$-polarized photon field shows that for the narrow bias window (the upper 
panel) all the current goes through the two spin components of the one-electron ground
state, $|\breve{3})$ and $|\breve{4})$. Even, for the broad bias, when the first photon replicas
of these two stats are in the bias window, the lower panel of Fig.\ \ref{SS-JLR-1G} indicates
that this is still to a large extent true, with the higher states contributing only slightly to the 
transport current. 

\begin{figure}[htb]
      \includegraphics[width=0.45\textwidth]{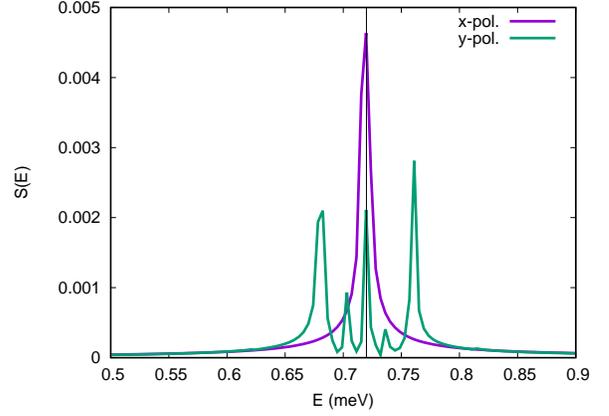}
      \caption{The spectral density $S(E)$ of the emitted cavity radiation of the system in its steady
               state for the electron-photon coupling $g_\mathrm{EM}=0.05$ meV, and a broad bias window
               with $\Delta\mu =0.86$ meV. 
               $V_g=2.0$ mV, $\hbar\omega =0.72$ meV (vertical black line), $\mu_L=1.96$ meV, $\mu_R=1.1$ meV, 
               and $\kappa = 1\times 10^{-3}$ meV.}
      \label{SS-fft-1G-bD}
\end{figure}

A close inspection of the partial currents for the broad bias window in the steady state
reveals that the partial currents from the central system into the right lead from the two
spin components of the one-electron ground state are slightly higher than the left
partial currents into these states indicating an electromagnetic transition active 
from higher states within the bias window. The spectral density of the emitted cavity
emission for the broad bias window, including the one-electron and the first photon 
replicas thereof, displayed in Fig.\ \ref{SS-fft-1G-bD} shows the same basic structure as
the spectral density for the narrow bias window in Fig.\ \ref{SS-fft-1G} with small additional
satellite peaks appearing for the $y$-polarized cavity field.

The situation for the $x$-polarized photon field (not displayed here)
shows the same picture as is reflected for the $y$-polarized photon field in
Fig.\ \ref{SS-JLR-1G}, regarding which states carry the transport current.

\subsection{Electroluminescence due to transport through the two-electron ground state}
It is more challenging to determine the vacuum contribution to the radiation for the transport through the
two-electron ground state since the current through it is low in the weak coupling 
serial tunneling regime for the leads and the central system, and as the first photon
replica of the two-electron ground state lies relatively high in the many-body energy
spectrum due to the Coulomb repulsion of the electrons.\cite{2016arXiv161109453G}

We select a bias window with the two-electron singlet ground state, $|\breve{6})$, 
just above $\mu_R$, by selecting the plunger-gate voltage $V_g=0.5$ mV, and $\mu_R=1.1$ meV.
A narrow bias window with only the two-electron state in it is defined with
$\mu_L=1.4$ meV, and thus $\Delta\mu =0.3$ meV. On the other hand, a broad bias
window with the first photon replica of the two-electron ground state  
within it in resonance with the first excitation of the two-electron ground state creating
the Rabi-split states, $|\breve{23})$ and $|\breve{24})$, by selecting
$\mu_L=3.4$ meV, and photon energy $\hbar\omega=2.0$ meV. The width of the resulting bias 
window is then $\Delta\mu =2.3$ meV.
 
\begin{figure}[htb]
      \includegraphics[width=0.45\textwidth]{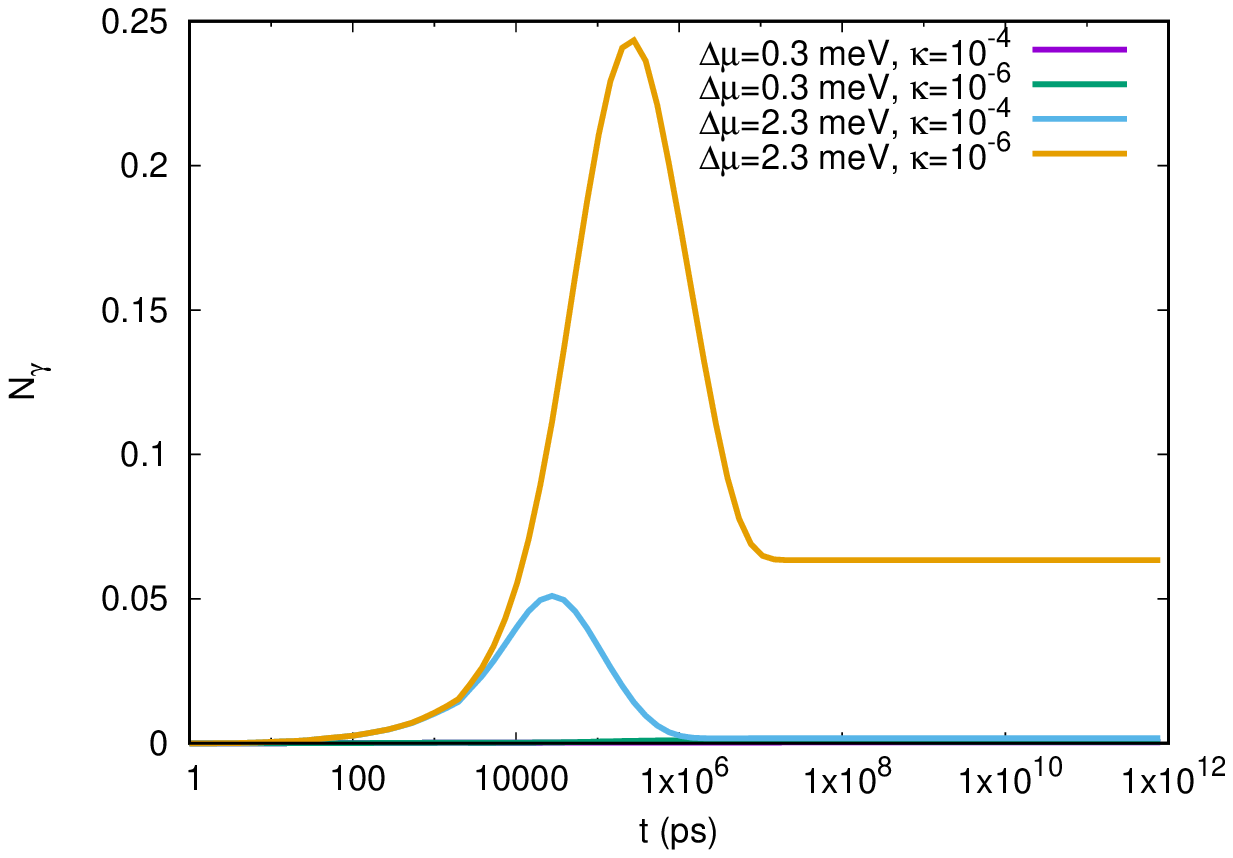}
      \includegraphics[width=0.45\textwidth]{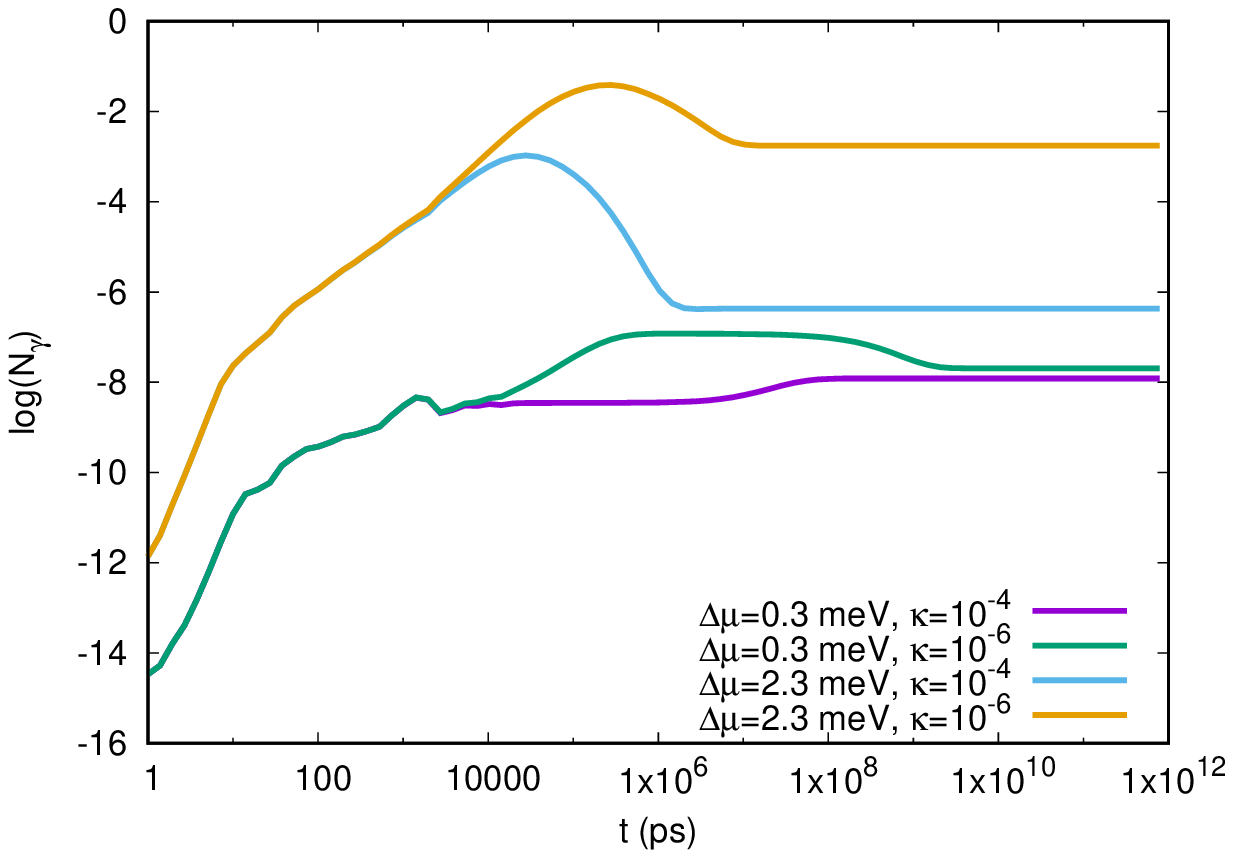}
      \caption{The mean photon number $N_\gamma$ as a function of time for the central system 
               initially in its vacuum state $|\breve{3})$ with $y$-polarized 
               cavity-photon field (upper panel), and on a logarithmic scale (lower panel). 
               $g_\mathrm{EM}=0.05$ meV, $V_g=0.5$ meV, $\hbar\omega =2.00$ meV,
               $\mu_R=1.10$ meV, and $\mu_L=1.4$ meV or $3.4$ meV. The coupling to the photon reservoir,
               the cavity decay constant $\kappa$ is measured in meV.}
      \label{Ng-2e-y}
\end{figure}

We start by analyzing the situation for a $y$-polarized cavity-photon field. 
The time evolution of the photon mean value, $N_\gamma$, is displayed in the upper panel of 
Fig.\ \ref{Ng-2e-y} and the same information is seen in the lower panel of Fig.\ \ref{Ng-2e-y},
but on a logarithmic scale.
We notice that it takes the system a longer time to reach the steady state, than for the
case of transport through the one-electron ground state.\cite{2016arXiv161109453G}
Furthermore, the transition to the steady state is marked by radiative processes, though 
to a lesser extent for the narrow bias window.\cite{Gudmundsson16:AdP_10} 
For the narrow bias window, $\Delta\mu =0.3$ meV, the photon mean value, $N_\gamma$,
is vanishingly small in the steady state, but for the broader bias window, $\Delta\mu =2.3$ meV,
a photon accumulation is seen for the slower cavity decay constant, $\kappa =1\times 10^{-6}$ meV.
The current through the central system is too low to cause a considerable accumulation of
photons for the higher decay constant, $\kappa =1\times 10^{-4}$ meV.    

\begin{figure}[htb]
      \includegraphics[width=0.45\textwidth]{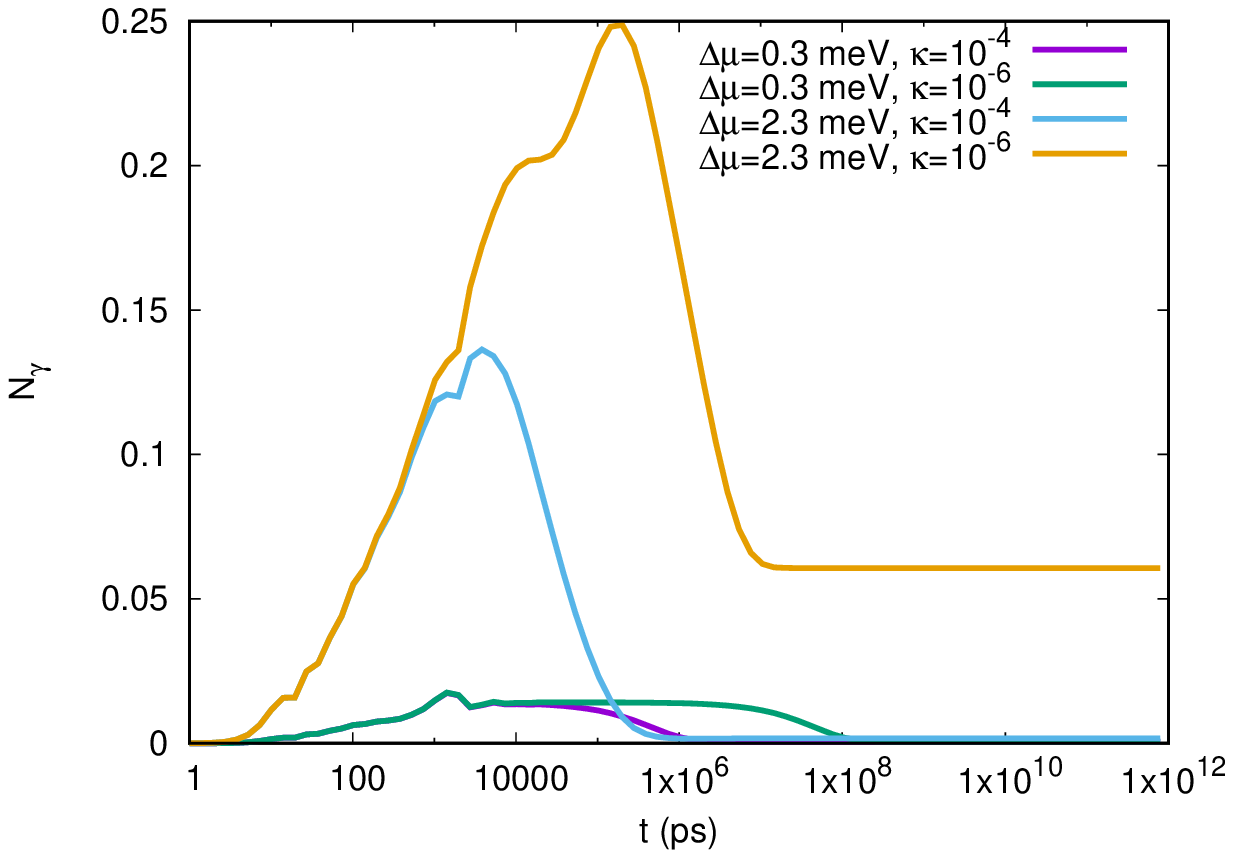}
      \includegraphics[width=0.45\textwidth]{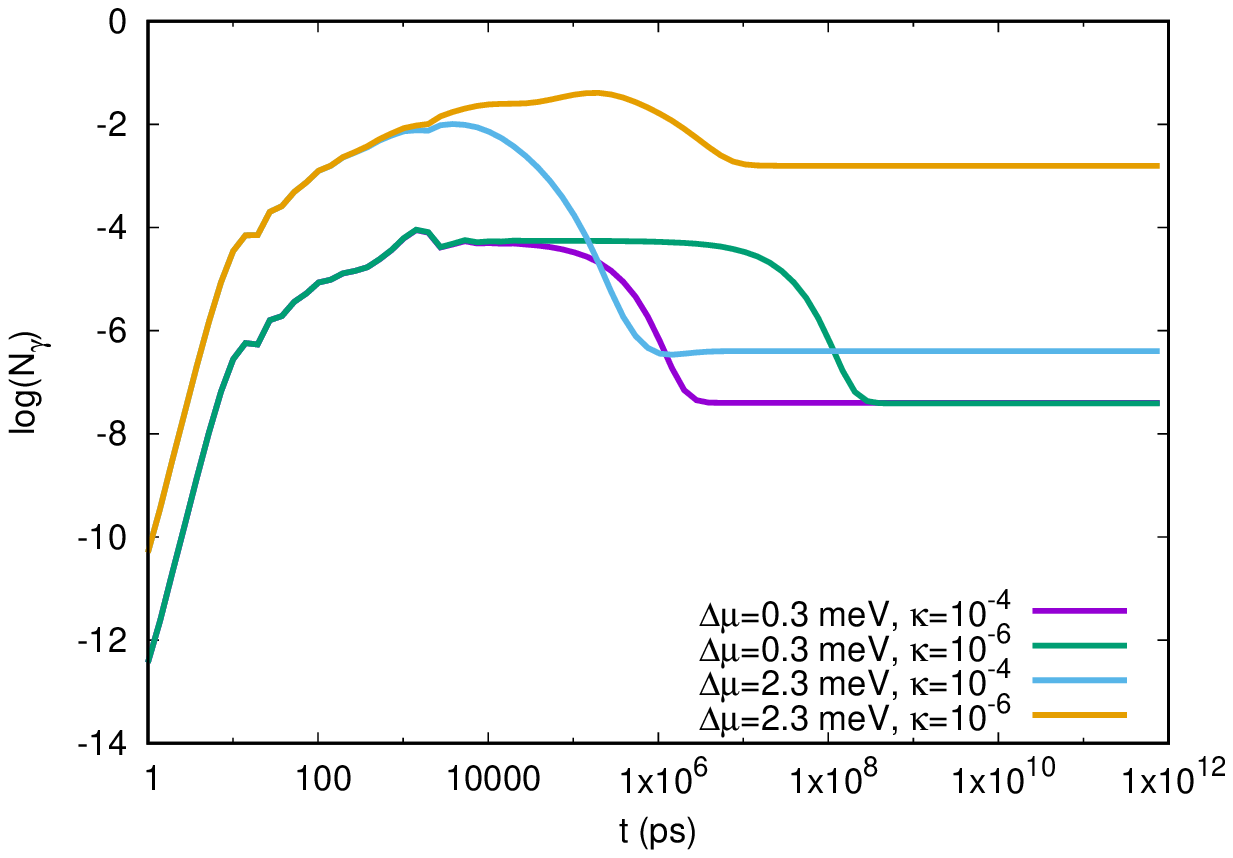}
      \caption{The mean photon number $N_\gamma$ as a function of time for the central system 
               initially in its vacuum state $|\breve{3})$ for $x$-polarized 
               cavity-photon field (upper panel), and on a logarithmic scale (lower panel). 
               $g_\mathrm{EM}=0.05$ meV, $V_g=0.5$ meV, $\hbar\omega =2.00$ meV,
               $\mu_R=1.10$ meV, and $\mu_L=1.40$ meV or $3.4$ meV. The coupling to the photon reservoir,
               the cavity decay constant $\kappa$ is measured in meV.}
      \label{Ng-2e-x}
\end{figure}

Similar results for the time evolution of $N_\gamma$ are obtained for the 
$x$-polarized cavity-photon field, seen in Fig.\ \ref{Ng-2e-x},
except for the fact that now the radiative relaxation processes bringing the system
close to the steady state start earlier and are stronger for the narrow bias window than
for the case of a $y$-polarized photon field. Again, a considerable accumulation of photons
is only attained for the slower cavity decay constant. 

\begin{figure}[htb]
      \includegraphics[width=0.45\textwidth]{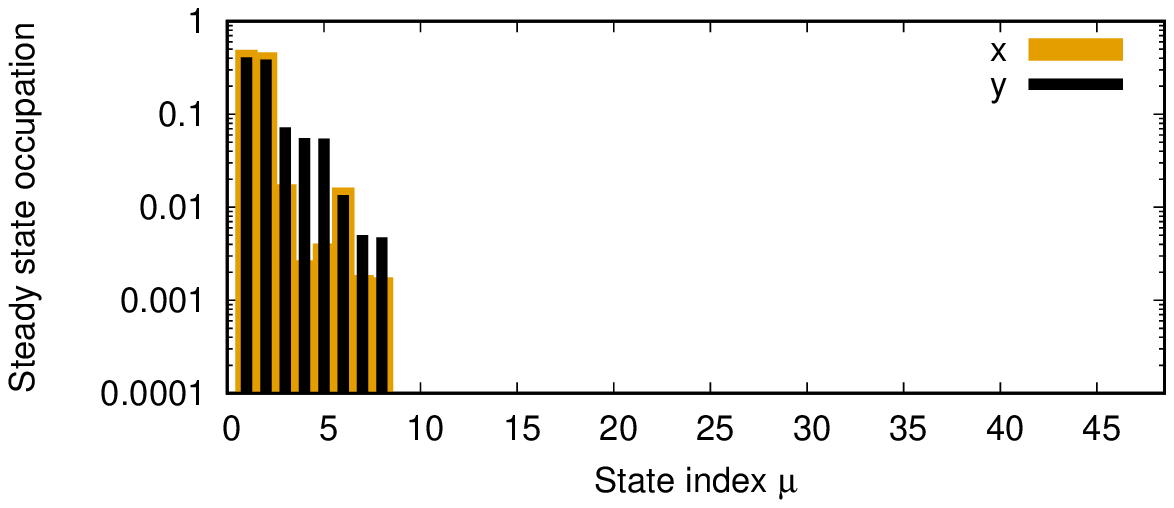}
      \includegraphics[width=0.45\textwidth]{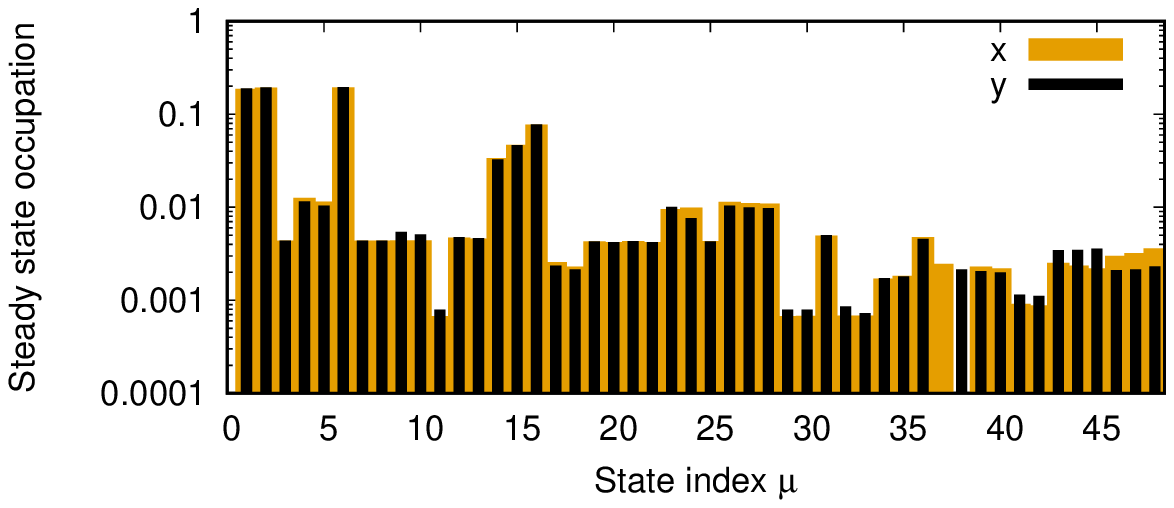}
      \caption{The steady-state occupation of the lowest 48 many-body states $|\breve{\mu})$
               for $V_g=0.5$ mV, $\mu_R=1.10$ meV, and $\mu_L=1.40$ meV (upper panel), and 
               $\mu_L=3.40$ meV (lower panel). The bias window includes states 
               $|\breve{6})$ - $|\breve{28})$. $g_\mathrm{EM}=0.05$ meV, 
               and $\kappa = 1\times 10^{-6}$ meV. The properties of the lowest 32 states 
               are presented in Fig.\ \ref{Fig-2e-structure}.}
      \label{SS-occ-2e}
\end{figure}

The steady-state statistical occupation of the many-body states of the central system
for the transport with the two-electron ground state in the bias window is presented 
in Fig.\ \ref{SS-occ-2e}. 
The narrow bias window, $\Delta\mu =0.3$ meV, only contains the two-electron ground state
$|\breve{6})$, but the broad bias window, $\Delta\mu =2.3$ meV, contains all states from
$|\breve{6})$ to $|\breve{28})$. We notice again that for the narrow bias window,
the upper panel of Fig.\ \ref{SS-occ-2e}, only states in the window, below it, and just above it
gain any occupation in the steady state. The two states just above it get occupied by thermal
smearing, as the temperature in the leads is 0.5 K. 
In contrast, for the broader bias window, the lower
panel of Fig.\ \ref{SS-occ-2e}, many states above the bias window are occupied. The occupation
is not very high due to the low current through the system, as was mentioned 
before.\cite{2016arXiv161109453G} We notice that the pure one-photon state with no electron 
component, $|\breve{11})$, has a nonvanishing occupation, though small, but the occupation
of the pure two-photon state, $|\breve{38})$ for $x$-polarization, and $|\breve{37})$
for the $y$-polarization, is too small to be visible on the scale used in the figure.
Closer inspection of the data indicates that the pure two-photon state $|\breve{38})$ 
for the $x$-polarization has a higher occupation of $1.3\times 10^{-5}$, while $|\breve{37})$
for the $y$-polarization is only one-tenth of that. This is in accordance with the fact that
the diamagnetic electron-photon interaction causes the Rabi-splitting for the case of
the $x$-polarization, since that interaction is composed of both one- and two-photon processes, while the
paramagnetic interaction in the lowest perturbation order only supports one-photon processes.  

The lower panel of Fig.\ \ref{SS-occ-2e} indicates that in the steady state the states
with highest occupation within the bias window are the two-electron states, including the
singlet ground state $|\breve{6})$, the lowest triplet states  $|\breve{14})$ -
$|\breve{16})$, and the first photon replicas of all these states.

Analysis of the partial current for the narrow bias window, $\Delta\mu =0.3$ meV,  with only the two-electron 
ground state ($V_g=2.0$ mV) within it shows only a very tiny current through the state ($|\breve{6})$) 
for $\kappa =10^{-6}$ meV, i.e.\ $I_6^{L,R}\approx 1.5\times 10^{-5}$ nA for the $x$-polarized photon field 
and $I_6^{L,R}\approx 1.3\times 10^{-5}$ nA for the $y$-polarized photon field at $g^\mathrm{EM}=0.05$ meV.
The mean value of photons in the two-electron ground state is $3.9\times 10^{-4}$ for the $x$-polarized
field and $3.2\times 10^{-4}$ for the $y$-polarized field, in accordance with the information displayed
in the lower panel of Figs.\ \ref{Ng-2e-y} and \ref{Ng-2e-x}. The generation of vacuum photons from the 
electron transport through the two-electron ground state is thus very small in our model. 

\begin{figure}[htb]
      \includegraphics[width=0.45\textwidth]{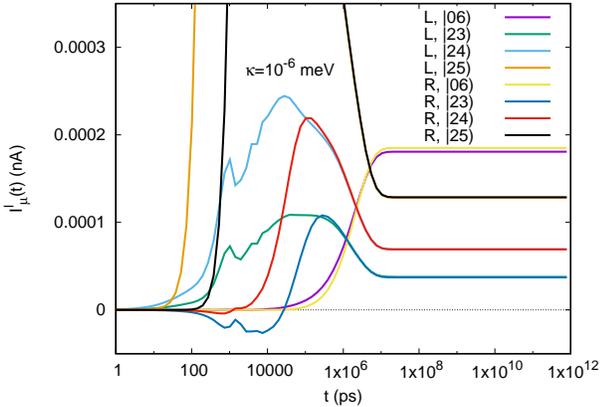}
      \caption{The partial current from the left lead (L) into state $|\breve{\mu})$, and the
               partial current from state $|\breve{\mu})$ into the right lead (R) as function
               of time broad bias window, $\Delta\mu = 2.3$ meV,
               panel. $V_g=0.5$ mV, $\hbar\omega =2.00$ meV, $g_\mathrm{EM}=0.05$ meV, 
               and $\kappa = 1\times 10^{-6}$ meV. $y$-polarized cavity photon field.} 
      \label{SS-JLR-2G}
\end{figure}

\begin{figure}[htb]
      \includegraphics[width=0.45\textwidth]{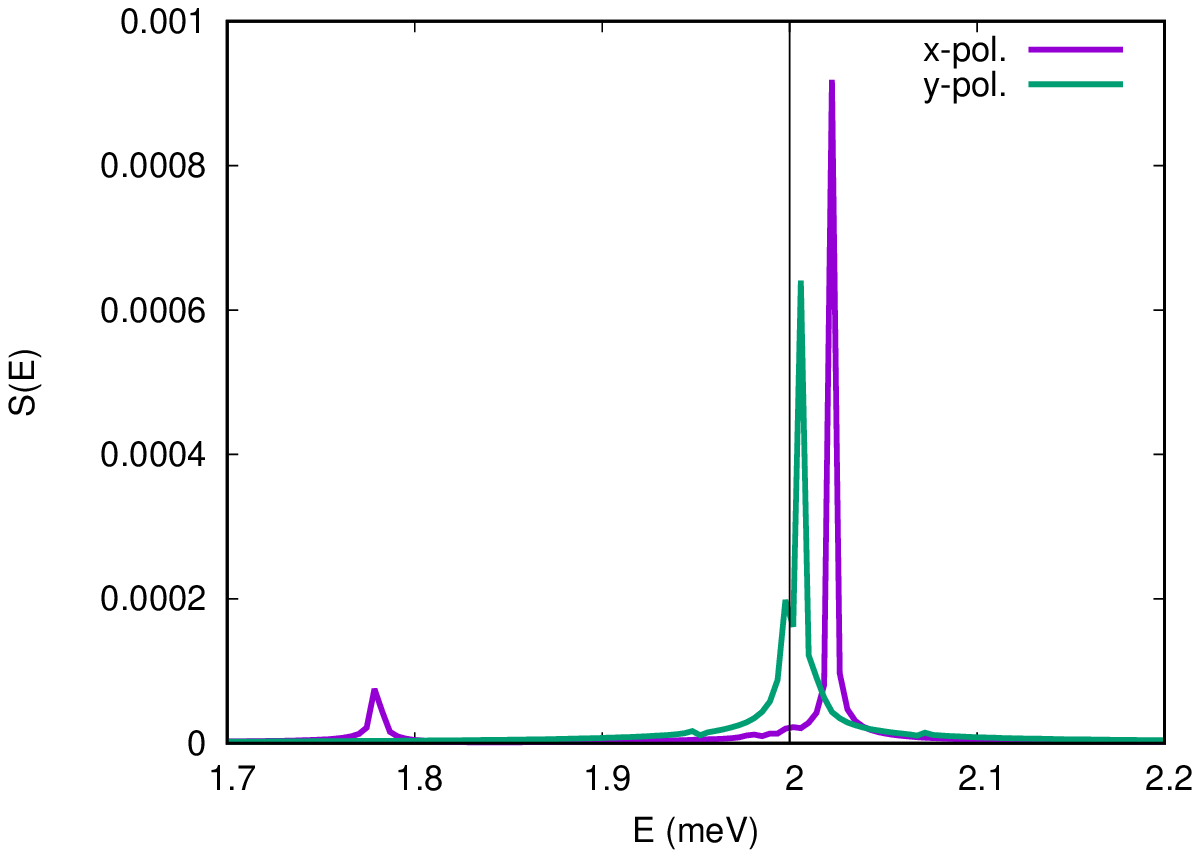}
      \includegraphics[width=0.45\textwidth]{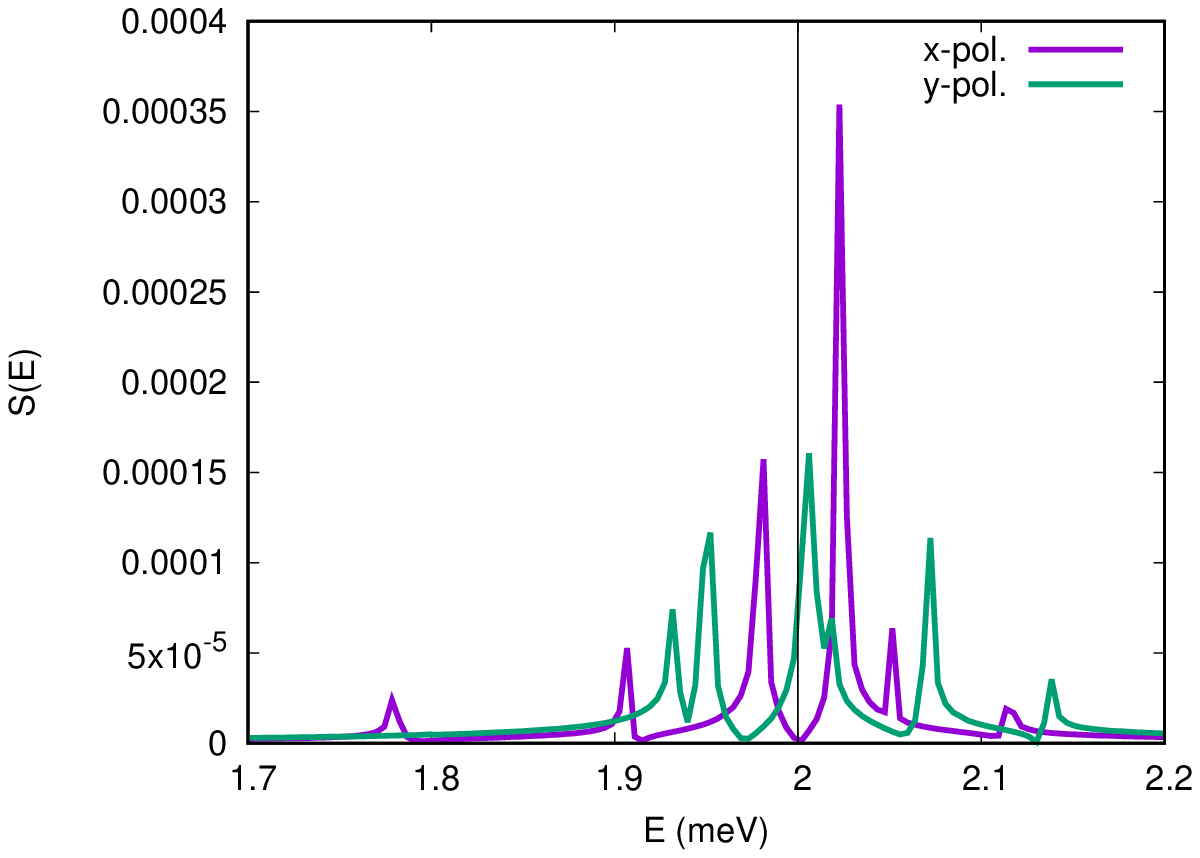}
      \caption{The spectral density $S(E)$ of the emitted cavity radiation of the system in its steady
               state for the narrow bias window $\Delta\mu =0.3$ meV (upper panel),
               and the broad bias window $\Delta\mu =2.3$  meV (lower panel). 
               $V_g=0.5$ mV, $\hbar\omega =2.00$ meV (vertical black line), $\mu_L=1.4$ meV, or $3.4$ meV, but
               $\mu_R=1.1$ meV, $\kappa = 1\times 10^{-4}$ meV, and $g_\mathrm{EM}=0.05$ meV.}
      \label{SS-fft-2G}
\end{figure}

The conventional electroluminescence of the two-electron states is enhanced by extending the bias window
to $\Delta\mu =2.3$ meV as is already evident from Figs.\ \ref{Ng-2e-y}-\ref{SS-occ-2e}. 
The partial currents are shown in Fig.\ \ref{SS-JLR-2G} for $\kappa =10^{-6}$ meV and an $y$-polarized
cavity field.
Notably, the current from the left lead and the current to the right lead from two-electron ground state,
$|\breve{6})$, are enhanced by photon active transitions from higher lying two-electron states
close to the bias in the left lead, $\mu_L =3.4$ meV. The spectral densities of the photon emission
are presented in Fig.\ \ref{SS-fft-2G}.
In case of the narrow bias window we notice that the low current through the two-electron
states for the $y$-polarized photon field hides the Mollow triplet. It can be recovered by changing the 
$y$-axis of the graph to a logarithmic scale.

\section{Discussion}
We have considered an anisotropic two-dimensional electron system of two parallel quantum dots
embedded in a short quantum wire placed in the middle of a rectangular photon cavity. 
We have shown that it is possible to select a Rabi-splitting either caused
by the para- or the diamagnetic electron-photon interactions by the choice of the cavity-photon
polarization. We find that both interactions lead to conventional,   
or ground-state electroluminescence (GSE), when current is driven through the system after coupling it to 
external leads with a large or small bias difference, respectively. 
Furthermore, we observe that even though the Rabi-splitting can be 
much smaller for the case of the $x$-polarization, enhancing the effects of the diamagnetic interaction, 
the electroluminescence is of similar order in both cases. 

As expected the electroluminescence is larger for the transport through the one-electron
ground state than for the transport through the two-electron ground state, due to the
low current through it in our sequential tunneling approach and the geometrical details
of the central system.\cite{2016arXiv161109453G}  In addition, the large photon energy
needed to couple the first photon replica and the first excitation of the two-electron
ground state into Rabi-split states includes many other states in the large bias-window 
needed to allow for the conventional electroluminescence.    

In terms of the normalized interaction strength, $\eta = \Omega_R/\omega$, measured by the
ratio of the Rabi- to the photon frequency, we find, and can quantify, vacuum effects 
before the traditional definition of the ultra-strong coupling regime with $\eta\approx 1$. 
This emphasizes the importance of considering high order effects in the electron-photon interaction, 
which we accomplish through the use of exact numerical diagonalization in a large basis, and leads
to a nonvanishing expectation value of the photon number operator in the one-electron
ground state well before $\eta\approx 1$. Second, special attention has to be paid to   
the geometry of a particular system within which the electron-cavity 
photon interactions are considered,\cite{ANDP:ANDP201500298} on one hand, and of the importance 
of the diamagnetic electron-photon interaction, on the other hand.\cite{PhysRevA.93.012120,PhysRevX.6.021014}

In atomic systems not placed in a photon cavity the diamagnetic electron-photon interactions 
only produces small corrections to most spectroscopic quantities measurable under 
{\lq\lq}normal{\rq\rq} conditions.\cite{PhysRevA.84.012510} Our model calculations indicate
that an anisotropic solid state system like double parallel quantum dots may be an ideal
experimental system to separate the effects of these two types of interactions if they are
placed in a photon cavity where the polarization of the fundamental mode can be chosen.

\begin{acknowledgments}
This work was financially supported by the Research Fund of the University of Iceland,
the Icelandic Research Fund, grant no.\ 163082-051, 
and the Icelandic Instruments Fund. HSG acknowledges support from Ministry of Science and 
Technology of Taiwan, under grant no.\ 103-2112-M-002-003-MY3.
\end{acknowledgments}

%
%
\bibliographystyle{apsrev4-1}
%

%
%
%
\end{document}